\documentclass[twocolumn,superscriptaddres,prb]{revtex4}

\usepackage{graphicx}
\usepackage{epsfig}
\usepackage{amsmath, amssymb}
\usepackage{verbatim}
\usepackage{multirow}

\begin{document}

\title{Classification of Abelian and Non-Abelian Multilayer \\ 
Fractional Quantum Hall States Through the Pattern of Zeros}
\author{Maissam Barkeshli}
\author{Xiao-Gang Wen}
\affiliation{Department of Physics, Massachusetts Institute of Technology,
Cambridge, MA 02139, USA }

\begin{abstract}

A large class of fractional quantum Hall (FQH) states can be classified 
according to their pattern of zeros, which describes the way ideal ground state wave functions go to zero as various 
clusters of electrons are brought together. In this paper we generalize this approach to classify multilayer
FQH states. Such a classification leads to the construction of a class of non-Abelian multilayer FQH states
that are closely related to $\hat{g}_k$ parafermion conformal field theories, where $\hat{g}_k$ is an affine simple Lie algebra. 
We discuss the possibility of some of the simplest of these non-Abelian states occuring in experiments 
on bilayer FQH systems at $\nu = 2/3$, $4/5$, $4/7$, etc. 

\end{abstract}

\maketitle

\section{Introduction}

One of the most important questions in condensed matter physics
relates to how we should characterize and classify the many different
quantum phases of matter. A large part of the story is the theory of
symmetry breaking, which tells us that we should classify various
phases based on the symmetries of the ground state wave function. Yet
with the discovery of the fractional quantum Hall (FQH)
states\cite{TS8259,L8395} came also the understanding that there are
many distinct and fascinating quantum phases of matter, called
topological phases, whose characterization has nothing at all to do
with symmetry. How should we characterize and systematically classify
the different possible topological phases that may occur in a FQH
system?

Let us run through a few obvious possibilities.  We know that the FQH
states contain topology-dependent and topologically stable degenerate
ground states, which allow us to introduce the concept of topological
order in FQH states.\cite{WN9077,W9505} Such topology-dependent
degenerate ground states suggest that the low energy theories
describing the FQH states are topological gauge field theories in 2+1
dimensions, called Chern-Simons
theories.\cite{BW9033,BW9045,R9002,WZ9290} So one possibility is that
we may try to classify the different FQH phases by classifying all of
the different possible Chern-Simons theories. But this is not a
satisfactory approach for non-Abelian FQH states because we do not
have a good way of knowing which Chern-Simons theories can possibly
correspond to a physical system made of electrons and which cannot. 

Another possibility comes through the connection between FQH
wavefunctions and conformal field theory (CFT). It was discovered
around 1990 that correlation functions in certain two-dimensional
conformal field theories may serve as good model wavefunctions for FQH
states.\cite{MR9162} Thus perhaps we may classify FQH states by
classifying all of the different conformal field theories that may be
used to construct a ``valid'' FQH wavefunction. While the connection
between FQH wave functions and CFT correlators has been extremely
fruitful in both constructing new model wavefunctions and
understanding their topological properties, the CFT approach is
incomplete in that there does not exist a complete classification of
all possible conformal field theories that may be used to construct
FQH wavefunctions. This precludes the possibility of systematically
classifying FQH states. The CFT approach also obscures the essential
physics and the essential properties of the conformal field theory
that allow one to obtain amenable FQH wavefunctions. 

In an attempt to obtain a systematic classification of FQH states, it
was shown recently that a wide class of single-component FQH states
and their topological excitations can be classified by their pattern
of zeros, which describe the way ideal FQH wavefunctions go to zero
when various clusters of particles are brought
together.\cite{WW0808,WW0809,BW0932} This analysis led to the
discovery of some new non-Abelian FQH states whose corresponding CFT
has not yet been identified. It also helped elucidate the role of CFT
in constructing FQH wavefunctions.  The CFT encodes the way the
wavefunction goes to zero as various clusters of electrons are brought
together. The order of these zeros must satisfy certain conditions and
the solutions to these conditions correspond to particular CFTs. Thus
in classifying and characterizing FQH states, one can bypass the CFT
altogether and proceed directly to classifying the different allowed
pattern of zeros and subsequently obtaining the topological properties
of the quasiparticles from the pattern of zeros. This construction can
then even be thought of as a classification of the allowed CFTs that
can be used to construct FQH states. Furthermore, these considerations
give way to a natural notion of which pattern of zeros solutions are
simpler than other ones. In this sense, then, one can see that the
Moore-Read Pfaffian quantum Hall state\cite{MR9162} is the
``simplest'' non-Abelian generalization of the Laughlin state. 

In this paper, we generalize the pattern-of-zeros classification to
multilayer FQH wave functions. Such a systematic classification leads
us to the construction of non-Abelian multilayer FQH states and also
helps identify the simplest non-Abelian
generalizations of the Halperin\cite{H8375} bilayer FQH states.
Therefore, in the same way that the Pfaffian FQH state is the simplest
non-Abelian single-layer state and so one of the first non-Abelian
states expected to be realized experimentally, our construction allows
us to identify the simplest non-Abelian bilayer states and
therefore some of the ones that deserve further consideration in
future experimental and numerical work.  

We would like to point out that the ``1D charge-density-wave''
characterization of single-component FQH states\cite{SL0604,BKW0608}
is closely related to the pattern-of-zeros approach.  Our
pattern-of-zeros approach for multilayer FQH states can also be viewed
as a way to generalize the ``1D charge-density-wave'' characterization
to multicomponent cases.\cite{SY0802}

This paper is organized as follows. We begin by describing the ideal
FQH wave functions that we can characterize by the pattern of zeros
and their relation to symmetric holomorphic polynomials.  In Section
\ref{pozCharacterization}, we describe the different ways to
characterize the pattern of zeros. In Section \ref{conditions} we find
the conditions that the pattern of zeros must satisfy in order to
describe valid FQH wave functions.  In Section
\ref{idealHamiltonians}, we sketch how one may begin to construct
ideal Hamiltonians whose ground states will be FQH wave functions with
a given pattern of zeros.  After a brief summary of the pattern of
zeros data and conditions in Section \ref{summary}, we explain in
Section \ref{cftSection} the relation between the pattern-of-zeros
approach and the CFT approach to FQH wave functions. In Section
\ref{Examples}, we describe some example solutions of this systematic
classification of multilayer FQH wave functions, which yields many
non-Abelian multilayer states. In Section \ref{Discussion} we discuss
some of the simplest of these non-Abelian bilayer states that may be
relevant for experiments on two-component quantum Hall systems and
that warrant further numerical study. 

\section{Fractional Quantum Hall States and Symmetric Polynomials }

The ground state wave function of a two-dimensional system of
electrons in the lowest Landau level can be written in the form
\begin{equation}
\Psi = \Phi(z_1, \cdots, z_{N_e}) e^{-\frac{1}{4} \sum_{i=1}^{N_e} |z_i|^2},
\end{equation}
where $z_i = x_i + iy_i$, $(x_i,y_i)$ are the coordinates 
of the $i^\text{th}$ electron, and $\Phi(z_1, \cdots, z_N)$ is a
holomorphic function of $z_i$. Since the electrons obey Fermi
statistics, $\Phi$ is anti-symmetric under interchange of any two
coordinates $z_i$ and $z_j$ when all of the electrons are identical.
In many physical situations, the electrons may be distinguished by
various quantum numbers, such as a spin index (when the Zeeman energy
is not too high), a layer index (in a multilayer two-dimensional
electron system), a valley index (such as in graphene and SiGe heterostructures), \it
etc \rm. In such cases, the ground state wave function in the lowest
Landau level may instead be written in the form
\begin{equation}
\Psi = \Phi(\{z_i^I \}) e^{-\frac{1}{4} \sum_{i,I} |z_i^I|^2},
\end{equation}
where $I = 1, \cdots, N_f$ is a flavor index and $N_f$ is the number
of different flavors. $\Phi$ is then necessarily antisymmetric only
under interchange of $z_i^I$ and $z_j^I$ for any $i$ and $j$.  Given
any antisymmetric polynomial $\Phi_{\text{anti-sym}}(\{z_i^I\})$, we
can uniquely construct a symmetric polynomial:
\begin{equation}
\Phi_{\text{sym}}(\{ z_i^I \} ) = \frac{ \Phi_{\text{anti-sym}}( \{z_i^I \}) }{\prod_{I; i<j} (z_i^I - z_j^I) }. 
\end{equation}
$\Phi_{\text{symm}}$ will also be a polynomial because
$\Phi_{\text{anti-sym}}$ must vanish when any two identical particles
approach each other. Thus the above division by the factor $\prod_{I;
i<j} (z_i^I - z_j^I)$ will never produce any poles in the resulting
function. 

Therefore to classify FQH phases of electrons, we can restrict our
attention mainly to symmetric polynomials $\Phi(\{z_i^I\})$, where
$\Phi$ is invariant under the interchange of $z_i^I$ and $z_j^I$ for
any $i$ and $j$, but not necessarily invariant under the interchange
of $z_i^I$ and $z_j^J$ if $I \neq J$.  In this paper we will often
refer to $I$ as a layer index. In the following, $\Phi$ will always
refer to such a symmetric multilayer polynomial. 

We will introduce data, such as $n$, $m$, and $S_{\vec a}$, to
characterize and classify bosonic FQH states (\ie symmetric
polynomials) $\Phi(\{z_i^I\})$.  From the above discussion, we see
that the same set of data also  characterizes fermionic FQH states
whose wave functions are given by $ \Phi_\text{anti-symm}(\{z_i^I\})=
\Phi(\{z_i^I\}) \prod_{I; i<j} (z_i^I - z_j^I) $.

\section{Ideal Hamiltonians and Ideal Wave functions}

Progress in understanding the various possible topological phases in
the FQH effect has occurred largely because of the discovery of
various kinds of model wave functions and the ideal Hamiltonians that
yield these wavefunctions as their ground states. In the pattern of
zeros approach, we classify all of the possible ideal wave functions
for FQH phases. In this section we will explain what we mean by
``ideal.'' 

For each topological phase in the FQH system, we want to have a
certain representative wave function that captures the topological
properties of the phase. The prototypical example of such an ideal
wavefunction is the Laughlin wave function at filling fraction $\nu =
1/m$:
\begin{equation}
\Phi_{1/m}( \{z_i \} ) = \prod_{i < j} (z_i - z_j)^m.
\end{equation}
At every particle in $\Phi_{1/m}$, there are $m$ zeros, and there are
no off-particle zeros. The exact ground state of the microscopic
Hamiltonian with the Coloumb interaction does not have this simple
property. Nevertheless, the ideal Laughlin wavefunction captures all of the
essential topological properties of these phases.

The Laughlin wavefunctions are also ground states of certain ideal
Hamiltonians that contain interaction potentials that are constructed
only from $\delta$-functions and their derivatives. For example, if
the interaction potential is of the form
\begin{equation}
V_1(z_1, z_2) = \delta(z_1 - z_2),
\end{equation}
then the following wavefunction is an exact ground state, with vanishing
total potential energy:
\begin{equation}
\Phi_{1/2} = \prod_{i < j} (z_i - z_j)^2.
\end{equation}
On the other hand, if the interaction potential between two electrons is of the form
\begin{equation}
V_2(z_1, z_2) = v_0 \delta(z_1 - z_2) + v_2 \partial^2_{z_1^*} \delta(z_1 - z_2) \partial^2_{z_1}, 
\end{equation}
with $v_0 > 0 $ and $v_2 > 0$, then the zero-energy ground state will be
\begin{equation}
\Phi_{1/4} = \prod_{i < j} (z_i - z_j)^4.
\end{equation}

More complicated ground states can be obtained through more complicated interactions.
For example, consider the following three-body interaction between electrons:
\begin{align}
V_{\text{Pf}} =& \mathcal{S} ( v_0 \delta(z_1 - z_2) \delta(z_2 - z_3) 
\nonumber \\
&- v_1 \delta(z_1 - z_2) \partial_{z_3^*} \delta(z_2 - z_3) \partial_{z_3} ),
\end{align}
where $\mathcal{S}$ is the total symmetrization operator between $z_1$, $z_2$, and $z_3$. 
An exact zero-energy ground state of this interaction is the Pfaffian wavefunction
at $\nu = 1$: 
\begin{equation}
\Phi_{\text{Pf}} = \mathcal{A}\left(\frac{1}{z_i - z_j}\right) \prod_{i < j} (z_i - z_j),
\end{equation}
where $\mathcal{A}$ is the total antisymmetrization operator between $z_1, \cdots, z_N$. 

Since the Pfaffian and Laughlin wavefunctions are exact zero-energy
ground states of interaction potentials that are constructed only from
$\delta$-functions and their derivatives, these wavefunctions can be
characterized completely by their pattern of zeros, i.e. by the order
of the zeros in the wavefunction as different numbers of particles are
taken together. These wave functions have certain ideal
properties; for example, the zeros that are bound to a particle lie precisely
at the location of that particle, not slightly away from it. In this
paper, we will classify ideal FQH wavefunctions, which are wave
functions that are exact zero energy ground states of such ideal
Hamiltonians. 

The existence of these ideal Hamiltonians is crucial. We can write
down any arbitrary complex function of $N$ variables, but we cannot
know whether it corresponds to a valid topological phase of matter in
the limit $N \rightarrow \infty$ unless we also know that it is the
ground state of some local, gapped Hamiltonian. If this is the case,
then it is conceivable that there exists some physical situation in
which the low energy effective interactions between electrons yields a
phase of matter that is in the same universality class as the ideal
wave functions that we constructed. Unfortunately, judging whether a
many-particle interacting Hamiltonian is gapped in the thermodynamic
limit is in general an intractable problem. So here we limit ourselves
to classifying ideal wave functions, which at least we believe can be
realized as ground states of local Hamiltonians.  Whether the
corresponding Hamiltonian can be gapped in the thermodynamic limit is
a question that we must attempt to answer along the way. 

The ideal quasihole wavefunctions for these states are also zero
energy eigenstates of their corresponding ideal Hamiltonians. The true
ground state is distinguished by being the unique translationally
invariant state with lowest angular momentum. Thus we can also
classify topologically distinct quasiholes by their pattern of zeros,
i.e. by the order of the zeros in the quasihole wavefunction as
different numbers of electrons are taken to the quasihole. While the
quasihole wavefunctions are also zero energy states, the
quasiparticles will be gapped and will have the same topological
properties as the quasiholes.  Therefore in the following, when we
discuss the pattern of zeros of the quasiparticles, we are referring
to the pattern of zeros of the quasihole wave functions. 

Not all FQH phases have such ideal wave functions. The hierarchy states
and the composite fermion states, for example, do not have ideal wave
functions. While these phases do have their own model wave
functions, they are not ideal in the sense that they cannot be
directly described by their pattern of zeros or written as a
correlation function of conformal primary operators in a CFT.
Therefore the pattern-of-zeros construction does not directly classify
these phases.  We will however discuss how they are related to the
ideal multilayer FQH wave functions in section \ref{sec:n11}.

\section{Pattern Of Zeros Characterization}
\label{pozCharacterization}

The spirit of the pattern-of-zeros approach is to consider bringing
together $a_I$ particles of type $I$, for $I = 1, \cdots, N_f$, and
asking how $\Phi$ goes to zero under such a procedure. The order of
the zero will be denoted $S_{\vec a}$, where $\vec a = (a_1, \cdots,
a_{N_f})$. In the following we will more precisely define $S_{\vec a}$
and discuss some different yet equivalent ways of characterizing the
pattern of zeros. This discussion is a straightforward generalization
of the discussion in the single-layer case.\cite{WW0808} 

\subsection{$S_{\vec a}$ characterization}

Consider a set of $a_I$ coordinates of each type $I$, and set $\vec a
= (a_1, \cdots, a_{N_f})$.  Define $S_{\vec a}$ as the minimal power
of $ ( \prod_{I = 1}^{N_f} \prod_{i=1}^{a_I} z_i^I )$ in the
polynomial $\Phi$. This means that if we set
\begin{align} 
z_i^I = \lambda \xi_i^I + z^{(\vec a)} &\;\;\;\;\;  i = 1, \cdots, a_I, \;\;\; \forall I,
\nonumber \\
z^{(\vec a)} = \frac{ \sum_{I=1}^{N_f}\sum_{i = 1}^{a_I} z_i^I }{\sum_I a_I} ,
&\;\;\;\;\;\;\;
\sum_{i, I} \xi_i^I = 0,
\end{align} and we take $\lambda \rightarrow 0$, then
\begin{equation}
\Phi \sim \lambda^{S_{\vec a}} P(\{\xi_i^I \}, z^{(\vec a)} , \{z_{a_I+1}^I, \cdots \}) + O(\lambda^{S_{\vec a}+1}),
\end{equation} where $P( \{\xi_i^I \}, z^{(\vec a)}, \{z_{a_I+1}^I,
\cdots \})$ is a polynomial in $\{\xi_i^I\}$ and the remaining
coordinates $z^{(\vec a)}$ and $\{z_{a_I + i}^I\}$. We refer to
$z^{(\vec a)}$ as the coordinate of an $\vec a$-cluster.  We assume
that $S_{\vec a}$ is independent of the choice of $z^{(\vec a)}$,
which must be the case for translationally invariant wavefunctions. We
also assume that $S_{\vec a}$ is independent of the choice of $\{
\xi_i^I \}$ and that different polynomials $P( \{\xi_i^I \}, z^{(\vec
a)}, \{z_{a_I+1}^I, \cdots \})$ obtained from different choices of
$\xi_i^I$ are linearly dependent. This is the assumption of \it unique
fusion \rm. 

We can immediately deduce some basic properties of $S_{\vec a}$. Since
$\Phi$ has no poles, it is clear that $S_{\vec a} \geq 0$. Since
$\Phi$ must be single-valued under rotating $\lambda$ in the complex
plane by an angle $2 \pi$, $S_{\vec a}$ must be an integer. Let
$S_{\vec e_I}$ be the minimal power of $z_1^I$; that is, $(\vec e_I)_J
= \delta_{IJ}$.  A translationally invariant $\Phi$ will have $S_{\vec
e_I} = 0$, otherwise it will vanish everywhere. 

Thus, for a translationally invariant polynomial, $S_{\vec a}$ is a
nonnegative integer that characterizes the order of zero that results
when the size of an $\vec a$-cluster goes to zero. 

\subsection{Derived Polynomials and the $D_{\vec{a} \vec{b}}$
characterization} 

In the previous section, we introduced the derived polynomials
$P(\{\xi_i^I \}, z^{(\vec a)} , \{z_{a_I+1}^I, \cdots \})$. As a
consequence of the unique fusion condition, these polynomials are
actually independent of $\{\xi_i^I \}$. We may consider more general
derived polynomials by bringing together other sets of coordinates in
$P$ to obtain $\tilde{P}(z^{(\vec a)}, z^{(\vec b)}, \cdots)$.  Then
we may consider bringing together an $\vec a$-cluster and a $\vec
b$-cluster:
\begin{align}
\tilde{P}(z^{(\vec a)}, &z^{(\vec b)}, \cdots)|_{z^{(\vec a)} \rightarrow z^{(\vec b)} \equiv z^{(\vec a + \vec b)}}
\nonumber \\
& \sim (z^{(\vec a)} - z^{(\vec b)})^{D_{\vec a \vec b}} \tilde{P}'(z^{(\vec a + \vec b)}, z^{(\vec c)}, \cdots)
\nonumber \\
& + O( (z^{(\vec a)} - z^{(\vec b)})^{D_{\vec a \vec b}+1}).
\end{align} Thus, $D_{\vec a \vec b}$ characterizes the order of the
zeros in the derived polynomials as a cluster of $\vec a$ electrons
are brought close to a cluster of $\vec b$ electrons.  The
unique-fusion condition assumes that the derived polynomials obtained
from different ways of fusion are always linearly dependent. 

The fact that $\Phi$ is a single-valued, symmetric polynomial implies
\begin{align}
D_{\vec a \vec b} = D_{\vec b \vec a} \in \mathbb{Z} ,
\;\;
D_{\vec a \vec a} = \text{even},
\;\;
D_{\vec a \vec b} \geq 0. 
\end{align}

We can deduce a relation between $D_{\vec a \vec b}$ and $S_{\vec a}$
as follows. The order of the zero obtained by creating an $(\vec a +
\vec b)$-cluster is $S_{\vec a + \vec b}$.  One way of creating such a
cluster is by first creating an $\vec a$-cluster, then creating a
$\vec b$-cluster, and finally bringing together the two clusters to
create an $(\vec a + \vec b)$-cluster. The order of zero in this case
will be $S_{\vec a} + S_{\vec b} + D_{\vec a \vec b}$. Thus 
$D_{\vec a \vec b}$ can be obtained from $\{ S_{\vec a} \}$
through the formula
\begin{equation}
D_{\vec a \vec b} = S_{\vec a + \vec b} - S_{\vec a} - S_{\vec b}.
\end{equation} Since $S_{\vec e_I} = 0$, where recall $(\vec e_I)_J =
\delta_{IJ}$ is the unit vector in the $I$ direction, we also have
\begin{equation}
S_{\vec a + \vec e_I} = S_{\vec a} + D_{\vec a, \vec e_I}.
\end{equation}
From this recursion relation and from the fact that $S_{\vec e_I} = 0$, we may
obtain $S_{\vec a}$ from the sequence $D_{\vec a \vec b}$. Therefore we may
equivalently label the pattern-of-zeros data using $\{ S_{\vec a} \}$
or $\{ D_{\vec a \vec b} \}$.

\subsection{Characterization by sequence of highest occupied orbitals}

The integer $S_{\vec a}$ has the following meaning. A polynomial
with $a_I$ particles of $I^\text{th}$ kind (\ie $a_I$ particles in the
$I^\text{th}$ layer) has a total order of $S_{\vec a}$.  In other
words the total angular momentum of the quantum Hall droplet is
$S_{\vec a}$ if the droplet has $a^I$ particles in the $I^\text{th}$
layer.  If we remove one particle from the $I^\text{th}$ layer, then
the total angular momentum of the quantum Hall droplet will be reduced
to $S_{\vec a -\vec e_I}$.  Thus we can interpret
\begin{equation}
\label{lSS}
 l^I_{\vec a}\equiv S_{\vec a}-S_{\vec a -\vec e_I}
\end{equation}
as the angular momentum of the highest occupied orbital in the $I^\text{th}$ layer
for a quantum Hall droplet with $a^J$ particles in the $J^\text{th}$ layer.
The $N_f$-dimensional sequence of vectors $\vec l_{\vec a} = (l^1_{\vec
a},...,l^{N_f}_{\vec a})$ will be called the sequence of highest occupied
orbitals (HOO).

We see that $l^I_{\vec a}$ makes sense only when $a^I >0$.  We will set
$l^I_{\vec a}=0$ when it does not make sense.  From \eq{lSS}, we also see that
there is one-to-one correspondence between the sequence $S_{\vec a}$ and $\vec
l_{\vec a}$. Thus we can also use $\vec l_{\vec a}$ to characterize the pattern
of zeros in the wave function.

\subsection{Relation to Angular Momentum on the Sphere}

A FQH wave function $\Phi( \{z_i^I\})$ defined on a sphere forms a
representation of $SU(2)$. In such a case, $z$ represents the
stereographic projection onto the plane of a point on the sphere. A
single particle in the lowest Landau level can fill any of the
$N_{\Phi} + 1$ orbitals, so the representation of $SU(2)$ formed in
this case is the one with angular momentum $J = N_{\Phi}/2$.  The
SU(2) Lie algebra is generated by 
\begin{align}
L^z = z \partial_z - J, \;\;L^- = \partial_z, \;\; L^+ = -z^2 \partial_z + 2Jz.
\end{align}
In the multilayer case, then, the angular momentum of a particle of
type $I$ will be $J_I = N_{\Phi}^I/2$ where $N_{\Phi}^I$ is the total
number of flux quanta through the sphere seen by the particles in the
$I^\text{th}$ layer. Note that here we allow the numbers of flux
quanta in different layers to be different.  The total angular
momentum of an $\vec a$ cluster in the $z$ direction will be the
eigenvalue of the operator
\begin{equation}
L^z_{\vec a} = \sum_I (\sum_{i = 1}^{a_I} z^I_i \partial_{z_i^I} - J_I). 
\end{equation}
The operator
$ \sum_I \sum_{i = 1}^{a_I} z^I_i \partial_{z_i^I}$
counts the total power of a polynomial.
Since the minimum total power of $\prod_I \prod_{i = 1}^{a_I} z_i^I$
is $S_{\vec a}$, the minimum total angular momentum of an $\vec a$-cluster
is given by $S_{\vec a} - \sum_I a_I J_I$. This means that the $\vec a$-cluster
carries an angular momentum of 
\begin{equation}
\label{Ja}
J_{\vec a} = 
\vec a \cdot \vec J - S_{\vec a}
=\frac12 \vec a \cdot \vec N_\Phi - S_{\vec a},
\end{equation}
where $\vec J=(J_1,...,J_{N_f})$ and
$\vec N_\Phi=(N_\Phi^1,...,N_\Phi^{N_f})$.
We will use this relation later to construct ideal Hamiltonians and to
place conditions on the pattern of zeros for when they can correspond
to rotationally invariant wave functions on the sphere. 

\section{Consistency Conditions}
\label{conditions}

For the pattern of zeros to describe a valid FQH wavefunction, it must
satisfy certain consistency conditions. We already encountered several
such conditions above. For instance, we found that $S_{\vec a}$ is a
nonnegative integer, 
$D_{\vec a \vec b} = S_{\vec a + \vec b} - S_{\vec a} - S_{\vec b}
\geq 0$, and $D_{\vec a \vec a} = S_{2 \vec a} - 2 S_{\vec a}$ is
even. In the following we find additional conditions that the pattern of
zeros must satisfy.

\subsection{Concave condition}

One of the most important conditions on the wave function is simply
the condition that the wave function have no poles.  This condition is
remarkably restrictive on the allowed pattern-of-zeros sequences.
Consider a derived polynomial $P(z^{(\vec a)}, z^{(\vec b)}, \cdots)$
and fix all coordinates but $z^{(\vec a)}$, thus viewing it as a
complex function $f( z^{(\vec a)})$. $f( z^{(\vec a)})$ has zeros at
isolated points, but no poles anywhere.  Some of the zeros are
located at $z^{(\vec b)}$, $z^{(\vec c)}$,  \etc. Those zeros are called
on-particle zeros.  The rest of the zeros are called off-particle zeros.  

If we imagine taking $z^{(\vec a)}$ around $z^{(\vec b)}$ without
enclosing any off-particle zeros, then $f$ will pick up a phase $2 \pi
D_{\vec a \vec b}$.  Similarly, if we take $z^{(\vec a)}$ around
$z^{(\vec c)}$ without enclosing any off-particle zeros, then $f$ will
pick up a phase $2 \pi D_{\vec a \vec c}$. Now consider taking
$z^{(\vec b)} \rightarrow z^{(\vec c)}$. Under such a process, some
nearby off-particle zeros will also be taken to $z^{(\vec c)}$.
Therefore, if we take $z^{(\vec a)}$ around a contour that encloses
both $z^{(\vec b)}$ and $z^{(\vec c)}$ in the limit that $z^{(\vec b)}
\rightarrow z^{(\vec c)}$, the complex function $f$ must change by a
phase that is greater than or equal to $2 \pi (D_{\vec a \vec b} +
D_{\vec a \vec c})$. The phase can never be less than this amount
because that would require the existence of off-particle poles that
get taken to $z^{(\vec c)}$ in order to diminish the strength of the
on-particle zeros. By definition, the phase change of $f$ under the
above procedure is $2 \pi D_{\vec a, \vec b + \vec c}$.  Therefore,
the condition that the wavefunction have no poles leads directly to
the following concavity condition on the integers $D_{\vec a \vec b}$:
\begin{equation}
D_{\vec a, \vec b + \vec c} \geq D_{\vec a \vec b} + D_{\vec a \vec c}.
\end{equation}
In cases where all of the zeros are located on the particles and there are no
off-particle zeros, the above inequality is saturated. This occurs
in the Laughlin states $\Phi = \prod_{i < j} (z_i - z_j)^m$, and their multilayer Abelian
generalizations, the Halperin states
\begin{equation}
\label{Halperin}
\Phi = \prod_{I;i < j} (z_i^I-z_j^I)^{K_{II}} \prod_{I < J; i,j} (z_i^I - z_j^J)^{K_{IJ}}.
\end{equation} 

In the following we will rewrite the concave condition as
\begin{align}
\Delta_3(\vec a, \vec b, \vec c) \geq 0,
\end{align}
\begin{align}
\Delta_3(\vec a, \vec b, \vec c) \equiv &D_{\vec a, \vec b + \vec c} - D_{\vec a \vec b} + D_{\vec a \vec c}
\nonumber\\
= & S_{\vec a + \vec b + \vec c} - S_{\vec a + \vec b} - S_{\vec a + \vec c} -  S_{\vec b + \vec c} 
\nonumber \\
& + S_{\vec a} + S_{\vec b} + S_{\vec c}.
\end{align}

\subsection{Cluster Condition}

The cluster condition is a way to associate some kind of grading to the polynomials
that is physically meaningful. Let $\{\vec n_I\}$ for $I = 1, \cdots, N_f$ 
be a set of vectors that generate an $N_f$-dimensional
lattice, where $N_f$ is, as before, the number of flavors of particles
(or the number of layers). The cluster condition 
states that the concave condition is saturated, \it i.e. \rm 
\begin{equation}
\label{clusterCond}
D_{\vec a, \vec b + \vec c} = D_{\vec a \vec b} + D_{\vec a \vec c},
\end{equation}
if either $\vec a$, $\vec b$, or $\vec c$ lie on the lattice generated by $\{ \vec n_I\}$. That is, if either
$\vec a$, $\vec b$, or $\vec c$ can be written as a linear combination with integer coefficients of the vectors
$\{ \vec n_I \}$. This means that a derived polynomial containing a 
$\vec k = \sum_I k_I \vec n_I$ cluster is non-zero unless $z^{(\vec k)}$ coincides
with the coordinates of another cluster; 
viewed as a function of the single variable  $z^{(\vec k)}$, it has no off-particle
zeros. A consequence of this is that if all of the particles are fused to form
$\vec n$-clusters, then the resulting derived polynomial has the Laughlin-Halperin
form (see (\ref{Halperin})) and there are no off-particle zeros.

The single-layer Read-Rezayi $Z_n$ parafermion wave functions satisfy 
an $n$-cluster condition and they are exact ground states of Hamiltonians 
with $n+1$-body interactions. For a fixed filling fraction, as
$n$ increases, the number of topologically distinct quasiparticles,
the ground state degeneracy on higher genus surfaces and the
complexity of interactions necessary to realize the state all
increase.  This suggests that the energy gap typically decreases with
increasing $n$.  Wave functions that do not obey a cluster condition
can be thought of as having infinite $n$ and are not expected to
correspond to gapped phases.  This intuition also comes from the
CFT approach to FQH wave functions; infinite $n$
corresponds to an irrational conformal field theory, which does not
yield a finite number of quasiparticles and a finite ground state
degeneracy on the torus. In the multilayer case, we may use the volume
of the unit cell spanned by $\{\vec n_I\}$ as one way to measure the
complexity of a given FQH state. 

The cluster condition is extremely powerful and simplifying because it allows us 
to determine the entire pattern-of-zeros sequence from knowledge of a ``small'' number of 
them. To see how this works, first observe using (\ref{clusterCond}) that
\begin{align}
D_{\vec n_J, \vec a} = \sum_I a_I D_{\vec n_J, \vec e_I} \equiv \sum_I m_{JI} a_I,
\end{align}
where we have defined the matrix $m_{JI} \equiv D_{\vec n_J, \vec e_I}$. So for any vector 
$\vec k = \sum_I k_I \vec{n}_I$, where $k_I$ is an integer and $\sum_I k_I (\vec{n_I})_J \geq 0$, we have: 
\begin{equation}
D_{\vec k, \vec a} = \sum_{IJ} k_I m_{IJ} \vec{a}_J.
\end{equation}
The above equations imply 
\begin{align}
D_{\vec n_I,\vec n_J} &= \sum_A n_{JA} D_{\vec n_I, \vec e_A} = \sum_A n_{JA} m_{IA} = (nm^T)_{JI}
\nonumber \\
&= \sum_A n_{IA} D_{\vec e_A, \vec n_J} = \sum_I n_{IA} m_{JA} = (m n^T)_{JI},
\end{align}
where we have also defined the matrix $n_{IJ} = (\vec n_I)_J$. 

In terms of the sequence $\{S_{\vec a}\}$, this implies that for
$\vec k = \sum_I k_I \vec{n}_I$, where $k_I$ is an integer,
\begin{align}
\label{SaCluster}
S_{\vec a + \vec k} &= S_{\vec a} + \sum_I k_I S_{\vec n_I} + \sum_{IJ} k_I m_{IJ} a_J 
\nonumber \\
 & + \frac{1}{2}\sum_{IJ} (nm^T)_{JI} (k_I k_J  - \delta_{IJ} k_I).
\end{align}
Therefore, all of the integers $S_{\vec a}$ are specified by the points $\vec{a}$ within the unit cell
spanned by $\{ \vec n_I \}$.

In terms of the HOO squence $\vec l_{\vec a}$, we have:
\begin{align}
\label{laCluster}
l_{\vec a + \vec k}^I &= S_{\vec a + \vec k} - S_{\vec a + \vec k - \vec e_I} 
\nonumber \\
 &= S_{\vec a} - S_{\vec a - \vec e_I} + \sum_A k_A m_{AI}
\nonumber \\
 &= l_{\vec a}^I + \sum_A k_A m_{AI}.
\end{align}
Finally, note that since $D_{\vec a, \vec a}$ is an even integer, we have:
\begin{align}
\text{even }  &= D_{\vec n_J \vec n_J} = \sum_I n_{JI} D_{\vec n_J, \vec e_I} = \sum_I n_{JI} m_{JI} 
\nonumber \\
 &= (nm^T)_{JJ}.
\end{align}

\subsection{Equal Area Layers}


The density profiles of the single-particle states in the lowest
Landau level, $z^m e^{- |z|^2/4l_B^2}$, are in the shape of a ring,
with a peak at a radius $r_m = \sqrt{2m} l_B$, where $l_B$ is the
magnetic length. Such a wave function has an angular momentum $m$.
When many of these orbitals are filled by particles,
the total wavefunction will describe a uniform, rotationally symmetric
state that goes to zero at a radius $r_{max} = \sqrt{2 m_{max}} l_B$,
where $m_{max}$ refers to the filled orbital with maximum angular
momentum. Therefore, a given quantum Hall wavefunction will describe a
QH droplet of area $4 \pi m_{max} l_B^2$, and $m_{max}$ is given by
the maximum power of $z_1$ (or $z_i$ for any other fixed $i$).
$m_{max}$ is also equal to the number of flux quanta, $N_{\Phi}$. 

An important constraint on the multilayer quantum Hall wavefunctions
is that they must describe systems in which each layer occupies the
same area, up to small corrections.  The requirement that each layer
occupies exactly the same area amounts to the requirement that each
layer has exactly the same number of flux quanta, $N_{\Phi}^I =
N_{\Phi}^J \equiv N_{\Phi}$. However, it is reasonable to include
states in which different layers occupy equal areas only up to
$O(N_e^0)$ corrections, where $N_e$ is the number of electrons. 

Such a requirement of approximately equal area layers is summarized in
the following equation:
\begin{align}
\label{equalAreaCond1}
\lim_{N_e \rightarrow \infty} \frac{N_{\Phi}^I}{N_{\Phi}^1} = 1.
\end{align}
We wish to see how this condition translates into a condition on the
pattern of zeros.  The conditions are slightly different depending on
whether we ultimately want to characterize gapped FQH phases of
fermions or bosons.  If we are interested in fermionic phases, we
require that $\Psi( \{z_i^I \}) =  \prod_{I; i< j} (z_i^I - z_j^I)
\Phi( \{ z_i^J \})$ be a valid FQH wave function of fermions, which
does not require that $\Phi( \{ z_i^I\})$ be a valid FQH wave function
of bosons.  In what follows we will explicitly analyze the bosonic
case, where we require $\Phi(\{z_i^I \})$ to be a valid multilayer FQH
wave function of bosons.

$N_{\Phi}^I$ is equal to the maximal power of $z_1^I$; for the boson
wavefunction $\Phi$, this is given by
\begin{align}
\label{NPhiPOZ}
N_{\Phi}^I = S_{\vec N } - S_{\vec N - \vec e_I},
\end{align}
where recall $N_I$ is the number of particles of type $I$. 
Using the cluster condition, we find
\begin{equation}
\label{NPhieqn1}
N_{\Phi}^J =   \sum_I N_I (n^{-1}m)_{IJ} + S_{\vec n_J} - S_{\vec n_J - \vec e_J}  - m_{JJ},
\end{equation}
where we have set $\vec N = \sum_I k_I \vec n_I$.
Requiring (\ref{equalAreaCond1}), we obtain the following condition on the pattern of zeros:
\begin{equation}
\label{equalAreaCond}
\sum_I (m^{-1}n)_{IJ} \geq 0.
\end{equation}
This can be seen most easily by ignoring the $O(N_e^0)$ terms in
(\ref{NPhieqn1}), taking $N_{\Phi}^I/N_{\Phi} \rightarrow 1$, and
inverting the result to obtain $N_I \sim N_{\Phi} \sum_I
(m^{-1}n)_{IJ}$, which must be nonnegative. From this analysis, we
learn that if $n^{-1}m$ is not invertible, then the pattern of zeros
cannot fix the ratio of particles $N_I/N_1$ in the different layers.
Therefore the corresponding FQH state has a gapless mode corresponding
to the relative density fluctuations between the different layers. 

As a simple example of this analysis, consider the $(1,1,1)$ Halperin
bilayer state, which is known to have a gapless density mode and for
which $n^{-1} m = \left( \begin{matrix} 1 & 1 \\ 1 & 1 \end{matrix}
\right)$, which is not invertible. A macroscopic number of particles
can freely go from one layer to the other without changing the area of
the quantum Hall droplets, signalling the existence of a gapless
relative density mode. 

Inverting (\ref{NPhieqn1}) yields 
\begin{equation}
\label{NPhieqn2}
\sum_{I} (m^{-1} n)_{IJ} (N_{\Phi}^I - S_{\vec n_I} + S_{\vec n_I - \vec e_I} + m_{II})= N_J,
\end{equation}
from which we can read off the filling fraction in each layer: 
\begin{equation}
\nu_I = \sum_J (m^{-1}n)_{IJ}.
\end{equation}
The total filling fraction is the sum of the filling fraction of each 
layer: $ \nu = \sum_I \nu_I$. 

For fermions, (\ref{NPhiPOZ}) is modified to 
\begin{equation}
N_{\Phi}^I = S_{\vec N} - S_{\vec N - \vec e_I} + N_I - 1,
\end{equation}
due to the extra factor $\prod_{I; i< j} (z_i^I - z_j^I)$ in
$\Psi(\{z_i^I\})$.  Note that $\{S_{\vec a} \}$ still describes the
pattern of zeros of the symmetric polynomial $\Phi$. The result for
fermions is therefore 
\begin{equation}
\nu_I = \sum_J (\mathbb{I} + n^{-1}m)^{-1}_{IJ} \geq 0,
\end{equation}
where $\mathbb{I}$ is the $N_f \times N_f$ identity matrix.  If
$(\mathbb{I} + n^{-1}m)$ is not invertible in the fermionic case, then
there are gapless relative density modes, which is why the filling
fraction in each layer becomes undefined. 

\subsection{Shift and Rotational Invariance on the Sphere}
\label{shiftRISphere}

Consider a multilayer quantum state with $N_I$ particles in the
$I^\text{th}$ layer. We want to put the quantum state on a sphere with
$N^I_\Phi$ flux quanta in the $I^\text{th}$ layer.  We would like to know
for which set of $N^I_\Phi$ can the quantum Hall state completely
fill the sphere? Naively, one may expect $N^I_\Phi$ and $N_I$ are
related by the filling fraction in each layer $N^I/\nu_I= N^I_\Phi$.
However the precise relation between the number of flux quanta and the
number of electrons includes a shift,
\begin{equation}
\label{shiftI}
\nu^{-1} \sum_I \nu_I N_{\Phi}^I = \nu^{-1} N_e - \mathcal{S},
\end{equation}
where $\mathcal{S}$ is of order 1 in the large $N_e$ limit (see eqn. (\ref{NPhieqn2}) ). 

More precisely, completely filling the sphere means that the quantum
Hall state is rotationally invariant with zero total angular momentum.
Using \eq{Ja}, we find that, for a bosonic FQH state characterized by
$S_{\vec a}$ with $N_I$ particles and $N^I_\phi$ flux quanta in the
$I^\text{th}$ layer, 
the maximum total angular momentum is given by
\begin{equation}
 J_{\vec N} = \frac12 \vec N \cdot \vec N_\Phi - S_{\vec N}
\end{equation}
where 
$\vec N=(N_1,...,N_{N_f})$ and
$\vec N_\Phi=(N^1_\Phi,...,N^{N_f}_\Phi)$.
For a fermionic FQH state characterized by
$S_{\vec a}$ with $N_I$ particles and $N^I_\phi$ flux quanta in the
$I^\text{th}$ layer, 
the maximum total angular momentum is given by
\begin{equation}
 J_{\vec N} = \frac12 \vec N \cdot \vec N_\Phi - S_{\vec N}
-\frac 12 \sum_I N_I(N_I - 1).
\end{equation}
In the above,
$\vec N_\Phi$ must satisfy
\begin{equation}
\label{SNSNeGEQ}
  \begin{array}{ll} 
N^I_\Phi \geq
S_{\vec N}-S_{\vec N-\vec e_I}
&\text{ for bosons} \\
N^I_\Phi \geq
S_{\vec N}-S_{\vec N-\vec e_I}+N_I-1
&\text{ for fermions}
  \end{array}
\end{equation}
in order for the wave function to fit into each layer.

Completely filling the sphere requires that
\begin{equation}
\label{SNSNe}
  \begin{array}{ll} 
N^I_\Phi=
S_{\vec N}-S_{\vec N-\vec e_I}
&\text{ for bosons} \\
N^I_\Phi=
S_{\vec N}-S_{\vec N-\vec e_I}+N_I-1
&\text{ for fermions}
  \end{array}
\end{equation}
and $J_{\vec N}=0$.
We see that $\vec N$ and $\vec N_\Phi$ must satisfy
\begin{equation}
\label{NNPhi}
\vec N \cdot \vec N_{\Phi} =
\left\{
  \begin{array}{ll} 
    2S_{\vec N} &\text{ for bosons} \\
    2S_{\vec N} + \sum_I N_I(N_I - 1) &\text{ for fermions}
  \end{array} \right.
\end{equation}
which implies (for both bosons and fermions)
\begin{align}
\label{NSS}
\sum_I N_I (S_{\vec N}-S_{\vec N-\vec e_I})=2S_{\vec N}.
\end{align}
If a given $\vec N$ does not satisfy \eq{NSS}, then the corresponding
quantum Hall state (with $N_I$ particles on the $I^\text{th}$ layer)
cannot completely fill the sphere.  For $\vec N$ that satisfies
\eq{NSS}, the corresponding quantum Hall state can completely fill the
sphere and has zero total angular momentum if $\vec N_\Phi$ is given
by \eq{SNSNe}. \eq{NSS} can generally be satisfied only if 
$\vec N$ lies on the lattice spanned by $\{ \vec n_I \}$. 

We would like to remark that it is easy to have different numbers of flux quanta
on different layers in numerical calculations.  The pattern of $(\vec
N,\vec N_\Phi)$ where the quantum Hall state has zero total angular
momentum on the sphere can be used as a fingerprint to identify different
quantum Hall states through numerical calculations (for examples, see Tables  
\ref{fluxNoTable23} and \ref{fluxNoTable45}).

\subsection{Additional constraints: $\Delta_3 =  \text{ even}$}
\label{delta3EvenSec}

The analysis of the single-layer case in \Ref{WW0808} has suggested an additional condition:
\begin{equation}
\label{delta3even}
\Delta_3(a, b, c) =  \text{even} .
\end{equation}
There, it was found that allowing $\Delta_3(a, b, c) =  \text{odd}$
allows for certain pattern-of-zeros sequences that either do not
correspond to single-valued wavefunctions (such as the square root of
the Pfaffian) or could not correspond to translationally invariant
wavefunctions. It was suggested that one way to rule out such
possibilities is to impose (\ref{delta3even}). How should this
condition be generalized to the multilayer situation?

One natural generalization is to impose $\Delta_3(\vec a, \vec b, \vec
c) =  \text{ even}$ for all $\vec a$, $\vec b$, and $\vec c$. However,
we find that this condition is too restrictive. It rules out certain
known FQH wavefunctions, such as the $su(3)_2/u(1)^2$
non-Abelian spin singlet states.\cite{AS9996, AR0149} The need to
relax this condition while still having it remain compatible with the
single-layer situation suggests that we should impose
(\ref{delta3even}) only for choices of $\vec a$, $\vec b$, $\vec c$
that are collinear through the origin. 

While it was found that allowing $\Delta_3(a, b, c) =  \text{ odd}$
allows for pattern-of-zeros sequences that do not seem to correspond
to valid translationally invariant, single-valued wavefunctions, there
are known cases of CFTs with $\Delta_3(a, b, c) =  \text{ odd}$ that
do seem to yield translationally invariant, single-valued
wavefunctions.  One such example is the so-called Gaffnian
wavefunction, which has $\Delta_3(1,1,1) = \text{ odd}$ and which can
be constructed using the minimal model CFT
$\mathcal{M}(5,3)$.\cite{SR0717} This CFT however is non-unitary. It
has been suggested that FQH wavefunctions constructed using
non-unitary CFTs correspond to gapless phases;\cite{R0908} whether
this is always necessarily the case is currently an important open
question in FQH theory.

These considerations suggest that in order to restrict ourselves to
pattern-of-zeros sequences that have a corresponding unitary CFT, we
should impose $\Delta_3(\vec a, \vec b, \vec c) =  \text{ even}$ for
those $\vec a$, $\vec b$, and $\vec c$ that are collinear through the
origin. In our search for pattern-of-zeros solutions, we will impose
this condition and analyze the resulting states. The precise
connection, if any, between this condition and valid FQH wavefunctions
that correspond to unitary CFTs remains to be clarified. 

\section{Ideal Hamiltonians} \label{idealHamiltonians}

Given a pattern-of-zeros sequence, it is important to be able to
construct a local, gapped Hamiltonian whose ground state wave function
has the given pattern of zeros. If this is possible, then we know that
the corresponding pattern-of-zeros sequence describes a topological
phase of matter. Whether this particular phase is realized in an
experiment then depends on the particular types of low energy
effective interactions between the electrons in the unfilled Landau
levels. 

We can go about constructing such a Hamiltonian by noticing that on a
sphere, the integers $S_{\vec a}$ are directly related to the angular
momentum of the $\vec a$-cluster. For an electron system on a sphere
with $N_{\phi}^I$ flux quanta for the $I$th layer, an electron of type
$I$ will carry an angular momentum $J^I = N_{\Phi}^I/2$. For an $\vec
a$-cluster, the maximum angular momentum is therefore $\vec a \cdot
\vec J$. However, for a polynomial $\Phi( \{ z_i^I \})$ described by a
pattern of zeros $\{ S_{\vec a} \}$, the maximum allowed angular
momentum of the $\vec a$ cluster is only $J_{\vec a} = \vec a \cdot
\vec J - S_{\vec a}$. The pattern of zeros forbids the appearance of
angular momentum $\vec a \cdot \vec J - S_{\vec a} + 1,\vec a \cdot
\vec J - S_{\vec a} + 2, \cdots, \vec a \cdot \vec J $ for any $\vec
a$-clusters in $\Phi( \{ z_i^i \})$.

Such a condition can be easily enforced by writing the Hamiltonian as
a sum of projection operators, $P_{S}^{(\vec a)}$. Let $P_{S}^{(\vec
a)}$ be a projection operator that acts on the $\vec a$-cluster
Hilbert space. $P_{S}^{(\vec a)}$ projects onto the subspace of $\vec
a$-clusters that have total angular momenta greater than $\vec a \cdot
\vec J - S$. Now consider the Hamiltonian
\begin{equation}
H_{\{S_{\vec a} \} } = \sum_{\vec a} \sum_{\vec a \text{-clusters}} P^{(\vec a)}_{S_{\vec a}},
\end{equation}
where $\sum_{\vec a \text{-clusters}}$ sums over all of the $\vec
a$-clusters for a fixed $\vec a$. The wave function described by
$\{S_{\vec a} \}$ will clearly be a zero-energy ground state of the
above $H_{\{ S_{\vec a} \}}$. In many cases, there is only one unique
ground state wave function with minimal total angular momentum, but in
general there can be many  independent polynomials with the same
pattern of zeros. In such a situation, the Hamiltonian would need to
be modified further to select for a particular polynomial with the
given pattern of zeros.  

In order for the above Hamiltonian to be local, $\sum_{\vec a}$ must
be limited to a small, finite number of $\vec a$-clusters.  But as a
result, we cannot guarantee in general that the ground state wave
functions will all be described by the sequence $\{ S_{\vec a} \}$ for
every $\vec a$, or even that they will obey the cluster condition. In
many of the known cases, such as the Laughlin, Moore-Read, and
Read-Rezayi wave functions, the sum over $\vec a$-clusters can indeed
be terminated after the first few clusters while still yielding a
unique zero energy ground state wave function with minimal angular
momentum which is described by $\{S_{\vec a} \}$. 

The above construction for $H_{ \{S_{\vec a} \}}$ should therefore be
viewed as a starting point for constructing an ideal Hamiltonian that
is local, gapped, and whose unique minimal angular momentum ground
state wave function is described by $\{S_{\vec a}\}$. In some of the
simplest cases, we know that this construction suffices.

\section{Summary: Pattern-of-Zeros Data and Conditions}
\label{summary}

We have found that the polynomials $\Phi(\{ z_i^I \})$, $I = 1,
\cdots, N_f$, that may correspond to stable FQH states are described
by the following data:
\begin{align}
n , \;\;\;\; m, \;\;\;\; \{S_{\vec a} \}, 
\end{align}
where $n$ and $m$ are $N_f \times N_f$ matrices with integer entries
that satisfy
\begin{align}
m_{IJ} \geq 0, \;\;\;\;\;\; n_{IJ} \geq 0,  \;\;\;\;\;\; \text{det } n \neq 0
\nonumber \\
m n^T = n m^T, \;\;\;\;\;\;  (m n^T)_{II} = \text{ even}.
\end{align}
The above implies that $n^{-1}m$ is a symmetric matrix.  Furthermore,
for the pattern of zeros to fix the relative densities of particles in
each layer, we have
\begin{align}
\label{mninv}
(n^{-1}m)_{IJ} &\ \ \text{ is invertible (for bosons)}
\nonumber\\
(n^{-1}m)_{IJ}+\del_{IJ} &\ \ \text{ is invertible (for fermions)}.
\end{align}
Otherwise, there are gapless relative density fluctuations. We also have
\begin{align}
\nu_I &\geq 0,
\nonumber \\
\nu_I &= \left\{  \begin{array}{ll}
    \sum_{J} (m^{-1}n)_{IJ} & \mbox{ for bosons} \\
    \sum_{J} (\mathbb{I} + n^{-1} m)^{-1}_{IJ} & \mbox{ for fermions } \\
  \end{array}\right.
\end{align}
The total filling fraction is $\nu = \sum_I \nu_I$.

Note that the $\{S_{\vec a}\}$ need to be specified only for values of
$\vec a$ that are contained in the unit cell spanned by $\{ \vec n_I
\}$, where $\vec n_I$ corresponds to the $I^\text{th}$ row of the
matrix $n$: $(\vec n_I)_J = n_{IJ}$.

Using the cluster condition, we can determine $S_{\vec a}$ for all $\vec a$
simply from $m$, the fact that the $S_{\vec e_I} = 0$, and from the
values of $S_{\vec b}$ for which $\vec b$ lies in the unit cell spanned by $\{ \vec n_I \}$:
\begin{align}
S_{\vec a + \vec k} &= S_{\vec a} + \sum_I k_I S_{\vec n_I} + \sum_{IJ} k_I m_{IJ} \vec{a}_J 
\nonumber \\
 & + \frac{1}{2}\sum_{IJ} (n m^T)_{IJ} (k_I k_J  - \delta_{IJ} k_I),
\end{align}
where $\vec k = \sum_I k_I \vec n_I$. $S_{\vec a}$ must satisfy:
\begin{align}
\Delta_2 (\vec a, \vec a) &= \text{ even},
\nonumber \\
\Delta_2(\vec a, \vec b) \geq 0, \;\;&\;\; \Delta_3(\vec a, \vec b, \vec c) \geq 0,
\end{align}
where
\begin{align}
\Delta_2(\vec a, \vec b) \equiv S_{\vec a + \vec b} - S_{\vec a} - S_{\vec b},
\nonumber 
\end{align}
\begin{align}
\Delta_3(\vec a, \vec b, \vec c) \equiv &S_{\vec a + \vec b + \vec c} - S_{\vec a + \vec b}
- S_{\vec a + \vec c} - S_{\vec b + \vec c} 
\nonumber \\
&+ S_{\vec a} + S_{\vec b} + S_{\vec c}.
\end{align}
Finally we impose
\begin{align}
\Delta_3(\vec a, \vec b, \vec c) = \text{ even}
\end{align}
for those $\vec a$, $\vec b$, and $\vec c$ that are collinear through the origin.

\subsection{Product of Symmetric Polynomials
and Primitive Solutions}
\label{CFTsec}

Consider two polynomials of $\vec{n}$-cluster form, $\Phi$ and
$\Phi'$, and consider their product: $\tilde{\Phi} = \Phi \Phi'$. The
pattern of zeros of $\tilde{\Phi}$ is the sum of the pattern of zeros
of $\Phi$ and $\Phi'$:
\begin{equation}
\tilde{D }_{\vec \alpha \vec \beta} = D_{\vec \alpha \vec \beta} + D_{\vec \alpha \vec \beta}' .
\end{equation}
Similarly, the data in terms of $m$ and $S_{\vec \alpha}$ are also
additive.  Note that all of the conditions that we impose on the
pattern-of-zeros are linear.  Thus the pattern of zeros of
$\tilde{\Phi}$ is also valid. However, notice that the condition for
filling the sphere is not linear. There may be two FQH wavefunctions
$\Phi$ and $\Phi'$ that can fill the sphere but whose product
$\tilde{\Phi}$ cannot fill the sphere. 
 
Thus, we can divide the pattern-of-zeros solutions into primitive and
non-primitive solutions. Primitive solutions are solutions that cannot
be written as a sum of two other solutions.

\section{ Relation to Conformal Field Theory } \label{cftSection}

The pattern-of-zeros approach is closely related to the conformal
field theory approach to constructing FQH wavefunctions. In the CFT
approach, the symmetric polynomial $\Phi( \{ z_i^I \}) $ that
describes a multilayer FQH state can be written as a correlation
function of a set of electron operators $V_{e;I}$ in a CFT:\cite{WWH9476}
\begin{equation}
\Phi( \{z_i^I \}) = \lim_{z_{\infty} \rightarrow \infty} z_{\infty}^{2 h_{\vec N} } \langle V( z_{\infty}) \prod_{i,I} V_{e;I}(z_i^I) \rangle.
\end{equation}
The operators $V_{e;I}$ are written in the form:
\begin{align}
V_{e;I}(z) = \psi_{\vec e_I}(z) e^{i \sum_{J} M_{IJ} \phi_J(z)},
\end{align}
where $e^{i \sum_{J} M_{IJ} \phi_J(z)}$ is a vertex operator in a
$U(1)^{N_f}$ CFT. It has scaling dimension $\sum_{J} M_{IJ}^2/2$.
$\psi_{\vec e_I}$ is a simple current operator; that is, it satisfies
the following fusion relation:
\begin{align}
\psi_{\vec a} \psi_{\vec b} = \psi_{\vec a + \vec b}.
\end{align}
This Abelian fusion rule is the CFT version of the unique-fusion condition. 
The cluster condition implies that $\psi$ satisfies
\begin{align}
\psi_{\vec n_I} \sim 1,
\end{align}
where $(\vec n_I)_J = n_{IJ}$. An $\vec a$-cluster of electrons will
be described by the operator 
\begin{align}
V_{\vec a} = \prod_I V_{e;\vec e_I}^{a_I} = \psi_{\vec a} e^{i \sum_{IJ} a_I M_{IJ} \phi_J(z)} .
\end{align}
Thus we see that the cluster condition implies that an $\vec n_I$
cluster is described by a vertex operator $e^{i \sum_{JK} n_{IJ}
M_{JK} \phi_K(z)}$. If all of the particles are grouped into
$n$-clusters, then the corresponding derived polynomial will be just a
correlation function of vertex operators in a $U(1)^{N_f}$ theory,
which will have no off-particle zeros and will be of the
Laughlin-Halperin form. 

Let us denote the scaling dimension of the operator $V_{\vec a}$ as
\begin{equation}
h_{\vec a} = h_{\vec a}^{\text{sc}} + h_{\vec a}^{\text{ga}},
\end{equation}
where $h_{\vec a}^{\text{sc}}$ is the scaling dimension of the simple
current $\psi_{\vec a}$ and $h_{\vec a}^{\text{ga}}$ is the scaling
dimension of the vertex operator. Notice that since $\psi_{\vec n_I}
\sim 1$, the simple current scaling dimensions satisfy $h_{\vec a +
\vec n_I}^{\text{sc}} = h_{\vec a}^{\text{sc}}$.  The scaling
dimension of the Gaussian part is given by
\begin{align}
\label{scalingDimGa}
h_{\vec a}^{\text{ga}} = \frac{1}{2} a_I (M M^T)_{IJ} a_J.
\end{align}
The pattern of zeros are related to the scaling dimensions through the relation
\begin{align}
\label{pozSC}
D_{\vec a, \vec b} &= h_{\vec a + \vec b} - h_{\vec a} - h_{\vec b}
\nonumber \\
&= S_{\vec a + \vec b} - S_{\vec a} - S_{\vec b}.
\end{align} 
This allows us to obtain the scaling dimensions from the pattern of zeros. 
Using the cluster condition, some algebra shows
that $M M^T = n^{-1} m$ and so (\ref{scalingDimGa}) becomes
\begin{align}
\label{scalingDimGA2}
h_{\vec a}^{\text{ga}} = \frac{\vec a^T n^{-1} m \vec a}{2}.
\end{align}
The scaling dimensions of the simple-current part can also be
determined from the pattern of zeros by using the fact that $h_{\vec
n_I} = \frac{ (n m^T)_{II} }{2}$, $h_{0} = 0$, and applying
(\ref{pozSC}) iteratively. This yields:
\begin{align}
\frac{ (nm^T)_{II}}{2} = \sum_{A} n_{IA} h_{\vec e_A} + S_{\vec n_I}.
\end{align}
Multiplying both sides by $n^{-1}$ gives
\begin{align}
h_{\vec e_A} = h_{\vec e_A}^{sc} + \frac{\vec e_A^T n^{-1} m \vec e_A}{2} = n^{-1}_{AI} ( \frac{ (nm^T)_{II}}{2} - S_{\vec n_I}).
\end{align}
In a similar manner, one can obtain
\begin{align}
\label{scalingDimSC}
h_{\vec a} &= \sum_I a_I (n^{-1})_{IJ} (\frac{(nm^T)_{JJ}}{2} - S_{\vec n_J}) + S_{\vec a}.
\nonumber \\
&= h_{\vec a}^{sc} + \frac{\vec a^T n^{-1} m \vec a}{2},
\end{align}
which determines $h_{\vec a}^{\text{sc}}$ in terms of the pattern-of-zeros. 

Note that the correlation function of the Gaussian part is, leaving
the background charge implicit, 
\begin{align}
\langle \prod_{I;i} &e^{\sum_J M_{IJ} \phi_J}(z_i^I)\rangle 
\nonumber \\
&=\prod_{I; i<j} (z_i^I - z_j^I)^{ (M M^T)_{II}} \prod_{I,J;i,j} (z_i^I - z_j^J)^{ (M M^T)_{IJ} } 
\nonumber \\
&= \prod_{I; i<j} (z_i^I - z_j^I)^{ (n^{-1} m)_{II}} \prod_{I<J;i,j} (z_i^I - z_j^J)^{ (n^{-1} m)_{IJ} }.
\end{align}
Thus the FQH wave function is of the form
\begin{align}
\Phi(\{z_i^I \}) = \Phi_{\text{sc}}( \{z_i^I\}) \Phi_{\text{ga}}( \{z_i^I\}),
\end{align}
where
\begin{align}
\label{phiGA}
\Phi_{\text{ga}}(&\{z_i^I \}) = 
\nonumber \\
& \prod_{I; i<j} (z_i^I - z_j^I)^{ (n^{-1} m)_{II}} \prod_{I<J;i,j} (z_i^I - z_j^J)^{ (n^{-1} m)_{IJ} }.
\end{align}
$\Phi_{sc}$ arises from the correlation function of the simple current sector and is the ``non-Abelian part'' of the 
wavefunction. 

In this manner, each pattern-of-zeros solution corresponds to the
current algebra of a rational CFT.  The connection between the pattern
of zeros approach and the CFT approach can be thought of in the
following way. The pattern of zeros describes the essential properties
of the CFT that yield valid FQH wavefunctions.  So in order to
classify ideal FQH wavefunctions, one can bypass the CFT altogether
and go directly to the heart of the matter: characterizing the allowed
pattern-of-zeros solutions. Furthermore, since each pattern-of-zeros
solution corrresponds to a CFT, the pattern-of-zeros classification
can be viewed as a classification of the allowed CFTs that can be used
to construct FQH wave functions.

In this formulation, the pattern of zeros classifies all those ideal
FQH wavefunctions that can be formulated as a correlation function of
\it conformal   primary \rm  fields. There are also many FQH
wavefunctions, such as the hierarchy states and the Jain series, that
cannot be written in this way. These wavefunctions are outside of the
pattern-of-zeros classification.  However they may be closely related
to the pattern-of-zeros construction. We comment on this connection
elsewhere.

\subsection{Alternate Labeling}

Using (\ref{scalingDimSC}), we can derive a formula for
$S_{\vec a}$ in terms of $h_{\vec a}^{sc}$, $n$, and $m$:
\begin{equation}
S_{\vec a} = h_{\vec a} - \sum_I a_I h_{\vec e_I}.
\end{equation} 
Thus there is a one-to-one correspondence between the simple-current
scaling dimensions and the sequence $\{S_{\vec a} \}$. This means that 
there is yet another way to label the pattern of zeros.  
Earlier, we found that one convenient labeling of the pattern of zeros 
is with an $N_f \times N_f$ matrix $n$, an $N_f \times N_f$ matrix $m$,
and the value of the non-negative integers $S_{\vec a}$ for $\vec a$ lying
inside the unit cell spanned by the rows of $n$. An alternative, equivalent
labeling of the pattern of zeros is by specifying the following data:
\begin{equation}
n, \;\;\; m,  \;\;\; \{h_{\vec a}^{sc}\},
\end{equation}
for $\vec a$ lying inside the unit cell spanned by $n$. This labeling
is convenient because it makes close contact with the corresponding
CFT description; if $h^{sc}_{\vec a} \neq 0$ for some $\vec a$, then
the CFT has a non-trivial simple-current structure and therefore
generally also has some form of non-Abelian statistics. 

\subsection{Relevant CFTs for multilayer FQH states: $\hat g_k/u(1)^r$ parafermions}

In the single-layer case, many of the pattern-of-zeros solutions were
found to be closely related to the $Z_k$ parafermion CFTs of
Zamalodchikov and Fateev.\cite{ZF8515} What are the relevant CFTs to
expect in the multilayer case? The answer is that some of the
corresponding CFTs in the multilayer case will be closely related to
the $\hat{g}_k/u(1)^r$ parafermion CFTs that were constructed
by Gepner, where $\hat{g}_k$ is a simple affine Lie algebra at level
$k$ and $r$ is the rank of the Lie algebra $g$.\cite{G8710} The case
$g = su(2)$ is equivalent to the $Z_k$ parafermion CFTs of
Zamalodchikov and Fateev. 

The simple-current algebra of the $\hat g_k/u(1)^r$
parafermion CFT has the following structure. For every element $\vec
\alpha$ of the $r$-dimensional root lattice of $g$, associate a simple
current operator $\psi_{\vec \alpha}$. The simple-currents will have
the fusion rules
\begin{align}
\psi_{\vec \alpha} \psi_{\vec \beta} = \psi_{\vec \alpha + \vec \beta}.
\end{align}
Furthermore, $\psi_{\vec \alpha} = \psi_{\vec \beta}$ if $\vec \alpha
- \vec \beta$ is an element of $k$ times the long root lattice of $g$.
The scaling dimension of $\psi_{\vec \alpha}$ is given by
\begin{align}
h_{\vec \alpha}^{sc} = -\frac{\vec \alpha^2}{2k} + n(\vec \alpha),
\end{align}
where $n(\vec \alpha)$ is an integer equal to the minimum number of
roots from which $\vec \alpha$ is composed. The inner product
$\vec{\alpha}^2 = (\vec \alpha, \vec \alpha)$ is defined with respect
to the quadratic form matrix of $g$.

If we are considering quantum Hall states with $N_f$ layers, then we
would expect to see the appearance of these parafermion CFTs with rank
$r \geq N_f$. Therefore in the bilayer case, one class of states that
we expect to see should be related to $\hat g_k/u(1)^2$
parafermion CFTS where $g$ is a simple Lie algebra of rank 2.  There
are only three simple Lie algebras of rank 2: $su(3)$, $so(5)$, and
$G_2$.  Of these, only $su(3)$ is simply laced, so the long root
lattice is the same as the root lattice.  This means that for the
pattern-of-zeros solutions that correspond to $n = \left(
\begin{matrix} k & 0 \\ 0 & k \\ \end{matrix} \right)$,  we expect to
see solutions that correspond to $su(3)_k/u(1)^2$ CFTs.  In
Appendix \ref{su3para} we will describe the $su(3)_2/u(1)^2$
CFT in more detail. 

The parafermion CFTs for $g = so(5)$ and $g = G_2$, on the other hand, are more
complicated because the long root lattice is different from the root
lattice. For example, $so(5)_k/u(1)^2$ CFTs will generically be relevant for
$n = \left( \begin{matrix} 2k & 0 \\ 0 & 2k \\ \end{matrix} \right)$
while $(G_2)_k/u(1)^2$ CFTs will generically be relevant for 
$n = \left( \begin{matrix} 3k & 0 \\ 0 & 3k \\ \end{matrix} \right)$.

\section{Examples of Pattern of Zeros Solutions}
\label{Examples}

In this section we examine explicitly several simple bilayer pattern
of zeros solutions.  We first fix the  cluster structure $n$ to have a
simple form.  Then we try to find all the solutions
$\{S_{\vec a}\}$ that satisfy the conditions listed in section
\ref{summary}.

The simplest non-Abelian states can be obtained 
from the simplest cluster structures 
$n = \left( \begin{matrix} 1 & 1 \\ 0 & 2 \\ \end{matrix} \right)$ and
$n = \left( \begin{matrix} 2 & 0 \\ 0 & 2 \\ \end{matrix} \right)$.
These are the simplest non-Abelian generalizations of the bilayer
Abelian Laughlin-Halperin states. 

Note that by definition the ordering of the rows in $n$ is arbitrary;
we choose it so that $n_{II} \neq 0$. Interchanging the layers yields
the same physical system but corresponds to interchanging $n_{11}$
with $n_{22}$ and $n_{12}$ with $n_{21}$, so two matrices $n$ and $n'$
that are related by such an interchange are regarded as equivalent.

To list the solutions for those simplest cases, we may
use some known CFTs to construct the appropriate simple-current
algebra that corresponds to the pattern-of-zeros solutions. Using this
known CFT, we can then write the wavefunction explicitly. However, the
wave function that we write down may not be unique in some cases;
there may be several independent polynomials that have the same
pattern of zeros. This corresponds to there being several distinct
CFTs whose simple-current algebra possesses the same pattern of zeros.
In the following examples, we will make this choice when necessary so
that we can explicitly write down a wave function with a certain
pattern of zeros. 

\subsection{ $n = \left( \protect\begin{matrix} 1 & 0 \\ 0 & 1
\protect\end{matrix} \right)$  }
\label{sec:n11}

These states are all Abelian and correspond to the Halperin $(m,n,l)$
states. The unit cell spanned by the $\vec n_I$ consists only of the
points $(1,0)$ and $(0,1)$; By translation invariance $S_{(1,0)} =
S_{(0,1)} = 0$. Thus these states are completely characterized by the
matrix $m$ and are of the form
\begin{align}
\Phi = \prod_{i<j} (z_i - z_j)^{m_{11}} 
\prod_{i<j} (w_i - w_j)^{m_{22}} \prod_{i,j} (z_i - w_j)^{m_{12}}.
\end{align}
The $m$ matrix here is exactly the $K$ matrix that describes all
Abelian FQH states.\cite{WZ9290} We also have $h_{\vec a}^{sc} = 0$
for all of these states -- the CFT has no simple-current part and
corresponds to a $U(1)^2$ Gaussian CFT. 

Although the single-layer hierarchy states, such as the $\nu=2/5$ hierarchy
state, do not have ideal single-layer wave functions, there are ideal
multilayer states that have the same topological orders as those
single-layer hierarchy states.  For example, the topological order in
the $\nu=2/5$ hierarchy state is described by the $K$-matrix $ K =
\left( \protect\begin{matrix} 3 & 2 \\ 2 & 3 \protect\end{matrix}
\right) $ (in the symmetric basis).\cite{WZ9290,W9505}
Such a topological order can be represented by
the ideal bilayer state with
$n = \left( \protect\begin{matrix} 1 & 0 \\ 0 & 1
\protect\end{matrix} \right)$ and
$m = \left( \protect\begin{matrix} 3 & 2 \\ 2 & 3
\protect\end{matrix} \right)$.
So although the pattern-of-zeros construction does not directly
classify those single-layer hierarchy states, their topological orders
can still be described by the pattern-of-zeros approach.

\subsection{ $n = \left( \protect\begin{matrix} 1 & 1 \\ 0 & 2
\protect\end{matrix} \right)$  }

This choice of $n$ requires that the electron operators in the CFT
must take the form
\begin{align}
V_{e1} = \psi e^{i \sum_J M_{1J} \phi_J},
\nonumber \\
V_{e2} = \psi e^{i \sum_J M_{2J} \phi_J},
\end{align}
where $\psi^2 = 1$ and $\psi$ has scaling dimension that is integer or
half-integer. This latter fact can be obtained from the condition
$\Delta_3( (1,0), (1,0) , (1,0) ) = \text{ even}$.  One general way of
constructing such a simple-current operator is by expressing it as a
set of Majorana fermions from several copies of the Ising CFT:
\begin{align}
\psi = \psi^{(1)} \cdots \psi^{(a)},
\end{align}
where $\psi^{(i)}$ is the Majorana fermion from the $i$th Ising CFT.
Such an operator has scaling dimension $h_{\psi}^{sc} = a/2$ and gives
rise to the following FQH wavefunction:
\begin{align}
\Phi( \{z_i, &w_i\}) = \text{Pf}\left( \frac{1}{x_i - x_j} \right)^a \times \Phi_{ga}.
\end{align}
$x_i$ represents the coordinates in both layers:
\begin{equation}
x_i \equiv \left\{ 
    \begin{array}{cc}
      z_i & 1 \leq i \leq N_1 \\
      w_{i-N_1} & N_1 < i \leq N_2 \\
    \end{array}\right.
\end{equation}
$\Phi_{ga}$ is defined in (\ref{phiGA}).  Note that the simple-current
algebra in this case implies $\psi_{(1,0)}^2 = \psi_{(0,1)}^2 = 1$.
That is, these states can all also be viewed as satisfying the cluster
condition for $n = \left( \begin{matrix} 2 & 0 \\ 0 & 2 \end{matrix}
\right)$, but with a different choice $m$. For every pattern-of-zeros
solution found here, there is an equivalent one for $n = \left(
\begin{matrix} 2 & 0 \\ 0 & 2 \end{matrix} \right)$. 

Later, we will list the solutions with the $n = \left( \begin{matrix}
2 & 0 \\ 0 & 2 \end{matrix} \right)$ cluster structure.  Some of those
solutions actually have $n = \left( \begin{matrix} 1 & 1 \\ 0 & 2
\end{matrix} \right)$.

\subsection{ $n = \left( \protect\begin{matrix} 1 & 1 \\ 0 & 1 \protect\end{matrix} \right)$  }

Here, the electron operators must be of the form
\begin{align}
V_{e1} &= \psi e^{i \sum_J M_{1J} \phi_J},
\nonumber \\
V_{e2} &= e^{i \sum_J M_{2J} \phi_J}.
\end{align}
The fact that $\vec n_1 = (1,1)$ however also forces $\psi = 1$. Thus
in fact the corresponding CFTs do not have a simple-current part; they
all correspond to a $U(1)^2$ Guassian CFT. All of these states are
therefore Abelian and correspond to the Halperin $(m,n,l)$ states.
Thus, all pattern-of-zeros solutions where $n = \left( \begin{matrix}
1 & 1 \\ 0 & 1 \end{matrix} \right)$ actually
also satisfy the cluster condition for 
$n = \left( \begin{matrix} 1 & 0 \\ 0 & 1 \end{matrix} \right)$.

Using the cluster condition and the fact that $S_{(1,0)} = S_{(0,1)} = 0$,
it is easy to see that the pattern of zeros for these states are completely
characterized by the matrix $m$ and therefore that these states are identical
to the Halperin states.

\subsection{ $n = \left( \protect\begin{matrix} 1 & 0 \\ 0 & 2 \protect\end{matrix} \right)$  }

In this case, the electron operators in the CFT must take the form
\begin{align}
V_{e1} &= e^{i \sum_J M_{1J} \phi_J},
\nonumber \\
V_{e2} &= \psi e^{i \sum_J M_{2J} \phi_J},
\end{align}
where $\psi^2 = 1$. 
From $\Delta_3( (0,1), (0,1), (0,1) ) = \text{ even}$, it follows that the
scaling dimension of $\psi$ is integer or half-integer. This means that we can in
general write it as a product of $a$ Majorana fermion operators from
$a$ independent copies of the Ising CFT. Thus the pattern-of-zeros
solution for this choice of $n$ includes only two classes of states:
the Abelian Halperin states and the following version of the Pfaffian
states:
\begin{align}
\Phi( \{z_i,&w_i\}) = \text{Pf}\left( \frac{1}{z_i - z_j} \right)^a \times \Phi_{ga}
\end{align}
Such a state spontaneously breaks the discrete
$Z_2$ symmetry associated with interchanging the two layers. 

\subsection{ $n = \left( \protect\begin{matrix} 2 & 0 \\ 0 & 2
\protect\end{matrix} \right)$ : non-Abelian bilayer states  }

For this choice of $n$, the corresponding CFTs have two simple current
operators, $\psi_{(0,1)}$ and $\psi_{(1,0)}$, which each square to the
identity: $\psi_{(1,0)}^2 \sim \psi_{(0,1)}^2 \sim 1$.  Thus there are
a total of three distinct primary currents: $\psi_{(1,0)}$,
$\psi_{(0,1)}$, and $\psi_{(1,1)} \equiv \psi_{(1,0)} \psi_{(0,1)}$.
The pattern of zeros can be fully specified by specifying the scaling
dimensions of these simple current operators, $h_{(1,0)}^{sc}$,
$h_{(0,1)}^{sc}$, and $h_{(1,1)}^{sc}$, and the matrix $m$, which
specifies the $U(1)^2$ part of the electron operators in the CFT. 

Applying $\Delta_3(\vec a, \vec a, \vec a) = \text{ even}$ for this
choice of $n$ shows that the simple-current scaling dimensions are all
integer or half integer: $2 h_{\vec a}^{sc} \in \mathbb{Z}$. One
general way of encoding these fusion relations and the associated
scaling dimensions is to write the electron operators in the CFT using
simple-current operators that consist of multiple copies of the
$su(3)_2/u(1)^2$ parafermion CFT. The most general choice for
the electron operators can be written in the form
\begin{align}
V_{e;1} &= \psi_{\alpha_1}^{(1)} \cdots \psi_{\alpha_a}^{(a)} e^{i \sum_J M_{1J} \phi_J},
\nonumber \\
V_{e;2} &= \psi_{\beta_1}^{(a-c)} \cdots \psi_{\beta_b}^{(a+b-c)} e^{i \sum_J M_{2J} \phi_J},
\end{align} where $\psi_{\alpha}^{(a)}$ is a simple current from the
$a$th copy of the $su(3)_2/u(1)^2$ parafermion CFT and
$\alpha_i$ and $\beta_i$ refer to either $(1,0)$, $(0,1)$, or $(1,1)$.
Some explicit forms for such correlators of simple-current operators in the
$su(3)_2/u(1)^2$ parafermion CFT were discussed in
\Ref{AR0149}. Computing these correlation functions provides one way -- not 
necessarily unique -- of constructing a wave function with the desired pattern of zeros.  

These pattern-of-zeros solutions can naturally be grouped into two distinct
classes. In the first class, $V_{e;1}$ and $V_{e;2}$ contain only one 
kind of simple-current, either $\psi_{(0,1)}$, $\psi_{(1,0)}$, or 
$\psi_{(1,1)}$, from each copy of the $su(3)_2/u(1)^2$ CFT. For example,
if $\psi_{(1,0)}^{(i)}$ appears in $V_{e;1}$, then $V_{e;2}$ cannot contain
$\psi_{(1,1)}^{(i)}$ or $\psi_{(0,1)}^{(i)}$. In such a situation, we can
think of $\psi_{\alpha_i}^{(i)}$ as being the Majorana fermion from the
Ising CFT. This means that these states can be written in terms of multiple copies 
of the Ising CFT .
This class of multilayer states can be written by choosing
\begin{align}
V_{e;1} &= \psi^{(1)} \cdots \psi^{(a)} e^{i \sum_J M_{1J} \phi_J},
\nonumber \\
V_{e;2} &= \psi^{(a-c)} \cdots \psi^{(a+b-c)} e^{i \sum_J M_{2J} \phi_J},
\end{align}
where $\psi^{(a)}$ is the Majorana fermion from the $a$th copy of the Ising CFT. 
The wave function for this class of states is therefore:
\begin{widetext}
\begin{align}
\Phi( \{z_i, w_i\}) = \text{Pf} \left( \frac{1}{z_i - z_j} \right)^{a-c} 
\text{Pf}\left( \frac{1}{w_i-w_j}\right)^{b-c} \text{Pf}\left( \frac{1}{x_i - x_j}\right)^{c} 
\prod_{i<j} (z_i - z_j)^{\frac{m_{11}}{2}} \prod_{i<j} (w_i - w_j)^{\frac{m_{22}}{2}} \prod_{i,j} (z_i - w_j)^{\frac{m_{12}}{2}}.
\end{align}
\end{widetext}
$x_i$ represents the coordinates in both layers:
\begin{equation}
x_i \equiv \left\{ 
    \begin{array}{cc}
      z_i & 1 \leq i \leq N_1 \\
      w_{i-N_1} & N_1 < i \leq N_2 \\
    \end{array}\right.
\end{equation}

This is the simplest generalization of the single-layer non-Abelian
states to a class of non-Abelian bilayer states: the interlayer
Pfaffian states. The simplest version of this, with $a = b = c$, is
closely related to (and identical to for certain choices of $m$)
the spin-charge separated non-Abelian spin singlet
wavefunction proposed in \Ref{AL0205}. 

The second class of states cannot be written in terms of multiple copies
of the Ising CFT -- the full $su(3)_2/u(1)^2$ parafermion CFT is necessary.
The first class of states, which can be written only using the Ising CFT,
have the property that their pattern of zeros satisfies
$\Delta_3(\vec a, \vec b, \vec c) = \text{ even}$ for all
choices of $\vec a$, $\vec b$, and $\vec c$. The second class of states,
for which the full $su(3)_2/u(1)^2$ CFT is necessary,
satisfies  $\Delta_3(\vec a, \vec b, \vec c) = \text{ odd}$ 
for certain choices of $\vec a$, $\vec b$, and $\vec c$. 

Let us compare the simple-current algebra of the $su(3)_2/u(1)^2$
parafermion CFT to what one would obtain using two copies of the Ising
CFT. If we used two copies of the Ising CFT, we could have
$\psi_{(1,0)} = \psi^{(1)}$ with scaling dimension $1/2$,
$\psi_{(0,1)} = \psi^{(2)}$ with scaling dimension $1/2$, and
$\psi_{(1,1)} \equiv \psi^{(1)} \psi^{(2)}$ with scaling dimension
$1$. This satisfies $\Delta_3( (1,0), (0,1), (1,1) ) = \text{ even}$.
On the other hand, in the $su(3)_2/u(1)^2$ parafermion CFT,
the only difference is that $\psi_{(1,1)}$ also has scaling dimension
$1/2$. Thus in this latter theory, two fermions combine to give
another fermion. This yields $\Delta_3( (1,0), (0,1), (1,1) ) = \text{
odd}$.  The fact that there are valid single-valued
translationally invariant FQH wavefunctions that arise from unitary
CFTs and that have $\Delta_3(\vec a, \vec b, \vec c) = \text{ odd}$
for certain choices of $\vec a$, $\vec b$, and $\vec c$ suggests (see
Section \ref{delta3EvenSec}) that we should impose $\Delta_3(\vec a,
\vec b, \vec c) = \text{ even}$ not in general but only if $\vec a$,
$\vec b$, and $\vec c$ are collinear through the origin. 

Let us examine the pattern of zeros for a few of the simplest examples
of these non-Abelian bilayer states. There is a fermionic $\nu = 2/3$
state with 
\begin{align}
\label{InterPfPOZ}
m &= \left( \begin{matrix} 2 & 2 \\ 2 & 2 \end{matrix} \right) \;\;
& \{ S_{(2,0)} = 0, S_{(1,1)} = 0, S_{(0,2)} = 0 \}
\nonumber \\
\nu &= 2/3 \;\;& \{ h^{sc}_{(1,0)} = \frac{1}{2}, h^{sc}_{(1,1)} = 0, 
h^{sc}_{(0,1)} = \frac{1}{2} \}.
\end{align}
This is the pattern of zeros for the interlayer Pfaffian state, which is of the form
\begin{align}
\label{InterPfwf}
\Psi( \{z_i, w_i\}) = \text{ Pf}\left(\frac{1}{x_i - x_j} \right) \Phi_{(2,2,1)} (\{z_i, w_i \}).
\end{align}
We use the notation 
\begin{align}
\Phi_{(\alpha,\beta,\gamma)} = \prod_{i<j} (z_i - z_j)^{\alpha} \prod_{i<j} (w_i - w_j)^{\beta} \prod_{i,j} (z_i - w_j)^{\gamma}.
\end{align}
There are also fermionic states at $\nu = 4/5$ and $\nu = 4/7$. These have the following pattern of zeros:
\begin{align}
\label{45POZ}
m &= \left( \begin{matrix} 2 & 1 \\ 1 & 2 \end{matrix} \right) \;\;& \{ S_{(2,0)} = 0, S_{(1,1)} = 0, S_{(0,2)} = 0 \}
\nonumber \\
\nu &= 4/5 \;\;& \{ h^{sc}_{(1,0)} = \frac{1}{2}, h^{sc}_{(1,1)} = \frac{1}{2}, h^{sc}_{(0,1)} = \frac{1}{2} \}.
\end{align}
\begin{align}
\label{47POZ}
m &= \left( \begin{matrix} 2 & 3 \\ 3 & 2 \end{matrix} \right) \;\;& \{ S_{(2,0)} = 0, S_{(1,1)} = 1, S_{(0,2)} = 0 \}
\nonumber \\
\nu &= 4/7 \;\;& \{  h^{sc}_{(1,0)} = \frac{1}{2}, h^{sc}_{(1,1)} = \frac{1}{2}, h^{sc}_{(0,1)} = \frac{1}{2} \}.
\end{align}
The state at $\nu = 4/7$ is the non-Abelian spin singlet state that was proposed in \Ref{AS9996}.

Note once again that the pattern of zeros $m$ and $\{S_{\vec a}\}$ refer to the pattern of zeros of the symmetric
polynomial, $\Phi = \frac{\Psi}{\prod_{I;i<j} (z_i^I - z_j^I)}$.

\section{ Discussion of results and relation to experiment}
\label{Discussion}

In single-layer quantum Hall samples, a quantum Hall plateau is seen at
$\nu = 5/2$, but not at $\nu = 1/2$. The reason is that 
even though in all of these cases there is a single half-filled Landau
level, the existence of the two filled extra Landau levels modifies the
effective interactions between the electrons in the unfilled level.
In the $\nu = 5/2$ case, numerical calculations suggest that the 
these effective interactions are modified in such a way that a
non-Abelian quantum Hall state may be realized. 

Experiments on multicomponent quantum Hall systems should be able 
to probe an even wider variety of regimes with distinct 
effective interactions. For example, for a two-component FQH system, we can 
study systems in which the spin degree of freedom is present, 
two-dimensional electron systems with two quantum wells, wide single-layer
systems in which the electrons spontaneously form a double-layer system due
to Coulomb repulsion, or systems in which there may be two valleys for the
free quasiparticle spectrum (such as in graphene or SiGe heterostructures), etc. 
In many of these cases, experimentalists can also tune to some extent the degree of correlation 
between the two components. For example, in double layer systems, application
of a parallel magnetic field can tune the tunneling and correlation between
the layers. There may also be some degree of tunability in the
relative densities between the two components in addition to being able 
to probe FQH states with different numbers of filled Landau levels. With
this greatly increased amount of variability and tunability in the effective
interactions between electrons in the unfilled Landau levels, it is 
possible that a non-Abelian state can be realized in a two-component 
quantum Hall system. 

Since the pattern of zeros provides a systematic classification and characterization of a 
wide variety of quantum Hall states, it provides us with a 
general sense of how all of the non-Abelian bilayer states are related and which
ones are simpler than other ones. Just as we know that the 
single-layer Pfaffian quantum Hall state is the simplest non-Abelian 
generalization of the Laughlin states, we can determine the simplest non-Abelian
generalization of the Halperin bilayer states and therefore single
out some of the possibilities that may be experimentally viable.  

In \Ref{BW09}, we have given an overview of some of the simplest non-Abelian 
bilayer states that we find and that occur at filling fractions at which 
experiments on two-component FQH systems have already observed incompressible 
states. Here we briefly summarize that discussion and supplement details
of the calculations of various topological properties of the candidate states. 

Experiments have so far observed FQH plateaus in two-component systems at 
$\nu = 2/3$, $4/5$, $4/7$, $4/9$, $6/5$, $6/7$, $1/4$, etc.
\cite{ES9010,EB9283,ME9428,MS9722,CY9822,SJ9147} 
In some cases, these plateaus have been observed in both bilayer and spin-unpolarized single-layer
systems, while in others, the plateau has only been observed in one of them. 
At all of these filling fractions, there exists also one (or several) candidate 
Abelian phase(s); in most cases, it is assumed that these plateaus are described 
by one of the Abelian phases. However, the pattern-of-zeros construction also
yields many simple non-Abelian states at these filling fractions. In some
situations, we expect the non-Abelian states to be good candidate states.

There are at least two dimensionless quantities that are important
determining factors for which FQH state is realized. The first
parameter is $\alpha \equiv V_{\text{inter}}/V_{\text{intra}}$, where
$V_{\text{inter}}$ is the potential for interlayer repulsion and
$V_{\text{intra}}$ is the potential for intralayer repulsion. The
second parameter is $\gamma \equiv t/V_{\text{intra}}$, where $t$ is
the interlayer hopping amplitude. In the limit $\alpha \sim 0$ and
$\gamma \sim 0$, the system will be a FQH state that consists of two
independent single-layer FQH states in each layer. In the limit
$\gamma \gg 1$ and $\alpha \sim 0$, a single-layer FQH state may be
observed. But if we keep $\gamma \sim 0$ and increase $\alpha$ from
$\al\sim 0$, then the FQH state formed by two independent single-layer
FQH states in each layer must undergo a phase transition into either a
compressible phase or a new incompressible state.  In the latter case,
an Abelian hierarchy state (such as a bilayer composite fermion state)
may form, which would in most cases be a state described by a
$4\times4$ or more complicated $K$-matrix and would have four or more
edge modes. The other possibility is that a non-Abelian two-component
state may form. The pattern-of-zeros construction yields non-Abelian
two-component wave functions that have zeros when particles from
different layers approach each other, indicating that they can
accomodate situations in which $\alpha \sim 1$. Additionally, these
states generally have less than $4$ edge modes; if we use the number
of edge modes as a measure of the complexity of the state, then the
non-Abelian states are simpler and may therefore be realized
experimentally. 

At $\nu = 2/3$, experiments on wide single quantum wells have observed
a phase transition from a bilayer to single-layer state while
experiments on single-layer systems have seen a phase transition from
a spin-polarized to a spin-unpolarized state. In the limit $\alpha
\sim 0 $ and $\gamma \sim 0$, the system should be in the $(3,3,0)$
state. As $\alpha$ is increased while $\gamma \sim 0$, one possibility
is the $(1,1,2)$ state. This wave function appears unphysical, because
it has higher order zeros as particles from different layers approach
each other than particles from within the same layer. Another wave
function, which has the same topological order as $(1,1,2)$, is a
spin-singlet composite fermion state.\cite{WD9353} There are two other
plausible non-Abelian states in this situation. One is the following
interlayer Pfaffian state (see (\ref{InterPfPOZ}) and
(\ref{InterPfwf}) ):
\begin{equation}
\label{NA23}
\Psi_{2/3}|_{\text{inter}} = \text{ Pf}\left(\frac{1}{x_i - x_j} \right) \Phi_{(2,2,1)} (\{z_i, w_i \}).
\end{equation}
The other is the following intralayer Pfaffian state
\begin{align}
\label{NA23a}
\Psi_{2/3|_\text{intra}} 
&= 
\text{ Pf}\left(\frac{1}{z_i - z_j} \right) 
\text{ Pf}\left(\frac{1}{w_i - w_j} \right) 
\Phi_{(2,2,1)} 
(\{z_i, w_i \}),
\end{align}
which has even higher order zeros as particles from different layers
approach each other.  $\Psi_{2/3}|_{\text{intra}}$ has a cluster
structure $n = \left( \begin{matrix} 1 & 1 \\ 0 & 2 \end{matrix}
\right)$ while $\Psi_{2/3}|_{\text{inter}}$ has a cluster structure $n
= \left( \begin{matrix} 2 & 0 \\ 0 & 2 \end{matrix} \right)$.  At
$\nu=2/3$ there are also two single-layer possibilities that may be
realized as $\gamma$ is increased. These are the particle-hole
conjugate of the $\nu = 1/3$ Laughlin state and the $Z_4$ parafermion
Read-Rezayi state. 

\begin{center}
\begin{table}
\begin{tabular}{|c|l|c|c|}
\hline
 $\nu$ &  Proposed States & Edge Modes & Shift $\mathcal{S}$ \\
\hline
\hline
\multirow{6}{*}{$2/3$} & $(3,3,0)$ & $2$ & $3$ \\
      & $(1,1,2)$ & $2$ &  $1$ \\  
      & $2/3|_\text{inter}$ (see eqn. (\ref{NA23}))& $2\frac12$ & $3$ \\
      & $2/3|_\text{intra}$ (see eqn. (\ref{NA23a}))& $3$ & $3$ \\
      & $Z_4$ parafermion & $3$ & 3 \\
      & P-H conjugate of $\nu=1/3$ & $1_R+1_L$ & 0 \\
\hline
\multirow{2}{*}{$4/5$} & $(2/5, 2/5 | 0)$ & 4 & 4 \\
& $su(3)_2/u(1)^2$ (see eqn. (\ref{su3states})) & $2\frac{6}{5}$ & 3 \\
& $(2/3, 2/3 | 1)$  & $2_R+2_L $  & 0 \\
\hline
\multirow{4}{*}{$4/7$} & $(2/7,2/7|0)$ & 4 & 2 \\     
& $su(3)_2/u(1)^2$ (see eqn. (\ref{su3states})) & $2\frac{6}{5}$ & 3 \\
& $(2/5, 2/5 | 1)$ & 4 & 4 \\
& $(2/3, 2/3 | 2)$ & $1_R+3_L $ & 0 \\
\hline
\multirow{4}{*}{$1/4$} & $(5,5,3)$ & 2 & 5 \\
      & $(7,7,1)$ & 2 & 7 \\
      & Inter-layer Pfaffian (see eqn. (\ref{NA14})) & $2\frac12$ & 7 \\
      & Single-layer Pfaffian & $1\frac12$ & 5 \\
\hline
\end{tabular}
\caption{ \label{proposed}
Proposed explanations for incompressible states at experimentally relevant 
filling fractions, $\nu = 2/3$, $4/5$, $4/7$, and $1/4$,
in two-component FQH systems. The bilayer composite fermion state $(\nu_1, \nu_2 | m)$ \cite{SJ0113}
refers to the state $\prod_{i,j} (z_i - w_j)^m \Phi_{\nu_1}( \{z_i\}) \Phi_{\nu_2} (\{w_i\})$, where $\Phi_{\nu}$
is a single layer composite fermion state at filling fraction $\nu$. For $(2/3,2/3|m)$, we have
taken the single layer $2/3$ state to be the particle-hole conjugate of the Laughlin state. $n_R + n_L$ 
indicates that there are $n_R$ right-moving edge modes and $n_L$ left-moving edge modes. 
See Appendix \ref{calculations} for details of how to calculate the number of edge modes and the shift $\mathcal{S}$. 
}
\end{table}
\end{center}

At $\nu = 4/5$, $4/7$, and $4/9$, we have the following 
non-Abelian states (see (\ref{45POZ}) and (\ref{47POZ}) ):
\begin{align}
\label{su3states}
\Psi_{4/5} &=  \Phi_{sc}( \{z_i, w_i\}) \Phi_{(2,2,\frac{1}{2})} (\{z_i, w_i\} ),
\nonumber \\
\Psi_{4/7} &=  \Phi_{sc}( \{z_i, w_i\}) \Phi_{(2,2,\frac{3}{2})} (\{z_i, w_i\} ), 
\nonumber \\
\Psi_{4/9} &=  \Phi_{sc}( \{z_i, w_i\}) \Phi_{(4,4,\frac{1}{2})} (\{z_i, w_i\} ),
\end{align}
where $\Phi_{sc} = \langle \prod_i \psi_{1}(z_i) \psi_{2} (w_i) \rangle$ is a correlation function
in the $su(3)_2/u(1)^2$ parafermion CFT. These states all have $2\frac{6}{5}$ edge modes. 

The other set of proposed Abelian states are the bilayer composite fermion states\cite{SJ0113} 
$(\nu_0, \nu_0 | m)$, which  refer to the wave function
\begin{equation}
\Phi_{(\nu_0,\nu_0|m)} = \prod_{i,j} (z_i - w_j)^m \Phi_{\nu_0}(\{z_i\}) \Phi_{\nu_0}(\{w_i\}).
\end{equation}
Here $\Phi_{\nu_0}(\{z_i\})$ is a single-layer FQH state at filling fraction $\nu_0$. 
These states have $4$ edge modes, indicating that they may be less stable than the
alternative non-Abelian possibilities. 

Recently, an incompressible state was found at $\nu = 1/4$ and it is unclear
what phase this corresponds to and even whether it is a single-layer or
double-layer phase.\cite{LP0874} Some possibilites that have recently been
considered\cite{PM0915} are the $(5,5,3)$ and $(7,7,1)$ Halperin states and the
$\nu = 1/4$ single-layer Pfaffian. The pattern-of-zeros construction yields
many other alternative possibilities, perhaps the most physical (and simplest)
of which is the following interlayer Pfaffian:
\begin{equation}
\label{NA14}
\Psi(\{z_i, w_i \}) = \text{ Pf}\left(\frac{1}{x_i - x_j} \right) \Phi_{(6,6,2)} (\{z_i, w_i \}).
\end{equation}

\begin{table}
\begin{tabular}{|c|c|c|}
\hline
\multicolumn{3}{|c|} {Quasiparticles for Interlayer Pfaffian at $\nu = 2/3$} \\
\hline
CFT Operator & Total Charge & Scaling Dimension \\
\hline
$V_{e1} = \psi e^{i \sqrt{3/2} \phi_+ + i \sqrt{\frac{1}{2}} \phi_-}$ & 1 & 3/2 \\
\hline
$V_{e2} = \psi e^{i \sqrt{3/2} \phi_+ - i \sqrt{\frac{1}{2}} \phi_-}$ & 1 & 3/2 \\
\hline
$e^{i \frac{2}{3} \sqrt{\frac{3}{2}} \phi_-} $ & 2/3 & 1/3 \\
\hline
$e^{i \frac{1}{3} \sqrt{ \frac{3}{2} } \phi_+ + i \frac{1}{\sqrt{2}} \phi_- }$ & 1/3 & 1/3 \\
\hline
$\sigma e^{i \frac{1}{3} \sqrt{ \frac{3}{2}} \phi_-}  $ & 1/3 & 7/48 \\
\hline
$\sigma e^{i \frac{2}{3} \sqrt{\frac{3}{2}} \phi_+ + i \sqrt{\frac{1}{2}} \varphi_-} $ & 2/3 & 31/48 \\
\hline
$\sigma e^{i \frac{1}{\sqrt{2}} \phi_-}$ & 0 & 5/16 \\
\hline
$\psi$ & 0 & 1/2 \\
\hline
$\psi e^{i \frac{2}{3} \sqrt{\frac{3}{2}} \phi_+} $ & 2/3 & 5/6 \\
\hline
$\psi e^{i \frac{1}{3} \sqrt{ \frac{3}{2} } \phi_+ + i \frac{1}{\sqrt{2}} \phi_- }$ & 1/3 & 5/6 \\
\hline
\end{tabular}
\caption{
\label{IPfQPOps}
Quasiparticle operators from the CFT for $\nu = 2/3$ Interlayer Pfaffian states. The
scalar boson fields $\phi_+$ and $\phi_-$ are related to the total and relative density fluctuations of 
the two layers, respectively. $\sigma$ is the spin field in the Ising CFT, which has scaling dimension
$1/16$.}
\end{table}

\begin{table}[t]
\begin{tabular}{|c|c|c|}
\hline
$\nu$ & Charge $q_{min}$ & Scaling Dimension $h$ \\
\hline
$2/3|_\text{inter}$  & 1/3 & $\frac{1}{16} + \frac{1}{12}+0$ \\
\hline
$2/3|_\text{intra}$  & 1/6 & $\frac{1}{16} + \frac{1}{48} + \frac{1}{16}$ \\
\hline
4/5 & 1/5 & $\frac{1}{10} + \frac{1}{40} + \frac{1}{24}$ \\
\hline
4/7 & 1/7 & $\frac{1}{10} + \frac{1}{56} + \frac{1}{8}$ \\
\hline
4/9 & 1/9 & $\frac{1}{10} + \frac{1}{72} + \frac{1}{56}$ \\
\hline
1/4 & 1/8 & $\frac{1}{16} + \frac{1}{32}+0$ \\
\hline
\end{tabular}
\caption{
\label{qpData}
Charge and scaling dimensions of the quasiparticle operators with
minimal nonzero total charge in the non-Abelian bilayer states discussed
here. In the scaling dimension, the first term comes from the non-Abelian
part, the second term comes from the total density fluctuations (the
$U(1)$ part), the third term comes from the relative density fluctuations
of the two layers (also the $U(1)$ part). 
} 
\end{table}

In Table \ref{proposed} we summarize some of the filling fractions at
which incompressible states have been experimentally observed in
two-component FQH systems. For each filling fraction we list some of
the proposed wave functions that may characterize the topological
order of those phases, the number of edge modes, and their respective
shifts on the sphere.  We list the qasiparticles, their electric
charges, and their scaling dimensions for the interlayer Pfaffian
state at $\nu = 2/3$ in Table \ref{IPfQPOps}.  In Table \ref{qpData},
we list the quasiparticles with the minimal electric charge and their
scaling dimensions $h$ for the non-Abelian FQH states discussed in
this paper [see eqns. (\ref{InterPfwf}), (\ref{45POZ}), (\ref{47POZ}),
(\ref{su3states}), (\ref{NA14})].  Those minimally charged quasiparticles
may dominate interedge tunneling and give rise to the following I-V
curve: $I \prop V^{4h-1}$ in the $T=0$ limit.

In summary, we find many simple non-Abelian bilayer states that occur
at experimentally observed filling fractions. For certain effective
interactions among the electrons in the unfilled Landau levels, these
states may be more favorable than their Abelian counterparts. In these
cases, the non-Abelian states have larger interlayer correlations and
therefore may be energetically more favorable in situations in which
the interlayer repulsion is comparable to the intralayer repulsion.

\subsection{Conditions on Filling the Sphere}

A useful tool for identifying FQH states in numerical studies of exact
diagonalization on finite systems on a sphere is to look at what values of the
shift, $\mathcal{S} = \nu^{-1} N_e - N_{\Phi}$, a ground state with
zero total angular momentum is found. This then limits the
possibilities of which topological phase is realized in the system to
those that have that particular value of the shift. Similarly, in such
numerical studies of multilayer systems, one can look for the
different sets $(N_1,\cdots, N_f; N_{\Phi}^1,\cdots, N_{\Phi}^{N_f})$
that yield a ground state with zero total angular momentum. Each
topological phase will have its own list of $(N_1,\cdots, N_f;
N_{\Phi}^1,\cdots, N_{\Phi}^{N_f})$ that let it fill the sphere;
analyzing this can be a useful way of determining which topological
phase is obtained numerically. In Section \ref{shiftRISphere}, we
found conditions that $\vec N$ and $\vec N_{\Phi}$ should satisfy for
the FQH state to fill the sphere. 

\begin{table}
\begin{tabular}{|c c| c c c c c|}
\hline
\multicolumn{7}{|c|} {$\nu=2/3$ Interlayer Pfaffian} \\
\hline
& &\multicolumn{5}{|c|} {$N_1$} \\
\multirow{7}{*}{$N_2$} & & 2 & 4 & 6 & 8 & 10\\
\hline
  & 2  & (3,3)  & (7,5)   & (11, 7)  & (15,9) & (19,11) \\
  & 4  & (5,7)  & (9,9)   & (13,11)  & (17,13) & (21,15) \\
  & 6  & (7,11) & (11,13) & (15,15)  & (19,17) & (23, 19)\\
  & 8  & (9,15) & (13,17) & (17,19)  & (21,21) & (25,23) \\
  & 10 & (11,19) & (15,21) & (19, 23) & (23,25) & (27,27) \\
\hline
\end{tabular}
\caption{\label{fluxNoTable23}
Values of $(N_{\Phi}^1, N_{\Phi}^2)$ that yield rotationally invariant states
on the sphere for various choices of $(N_1, N_2)$ for the $\nu = 2/3$ interlayer Pfaffian.}
\end{table}
\begin{table}
\begin{tabular}{|c c| c c c c c|}
\hline
\multicolumn{7}{|c|} {$\nu=4/5$ $su(3)_2/u(1)^2$ parafermion state} \\
\hline
& &\multicolumn{5}{|c|} {$N_1$} \\
\multirow{7}{*}{$N_2$} & & 2 & 4 & 6 & 8 & 10\\ 
\hline
  & 2  & (2,2)  & (6,3)   & (10, 4)  & (14,5) & (18,6) \\
  & 4  & (3,6)  & (7,7)   & (11,8)  & (15,9) & (19,10) \\
  & 6  & (4,10) & (8,11) & (12,12)  & (6, 13) & (20, 14)\\
  & 8  & (5,14) & (9,15) & (13,6)  & (17,17) & (21,18) \\
  & 10 & (6,18) & (10,19) & (14, 20) & (18,21) & (22,22) \\
\hline
\end{tabular}
\caption{\label{fluxNoTable45}
Values of $(N_{\Phi}^1, N_{\Phi}^2)$ that yield rotationally invariant states
on the sphere for various choices of $(N_1, N_2)$ for the $\nu = 4/5$ $su(3)_2/u(1)^2$
parafermion state.}
\end{table}

For states that have a cluster structure $n = \left( \begin{matrix} 2
& 0 \\ 0 & 2 \end{matrix} \right)$, we find that the condition
(\ref{NSS}) becomes trivial as long as $\vec N = \sum_I k_I \vec n_I$,
where $k_I$ is an integer.  This means that as long as $N_1$ and $N_2$
are even and $N_{\Phi}^1$, $N_{\Phi}^2$ satisfy (\ref{SNSNe}), then
these states can fill the sphere.  In this case, we find that
(\ref{SNSNe}) reduces to the form \begin{align} \left( \begin{matrix}
N_{\Phi}^1 + \mathcal{S} \\ N_{\Phi}^2 + \mathcal{S} \end{matrix}
\right) = M  \left(\begin{matrix} N_1 \\ N_2 \end{matrix} \right),
\end{align} where $M$ is a $2\times 2$ matrix and $\mathcal{S}$ is the
shift, which can be calculated using eqn. (\ref{shift}). The states
that we have been considering are of the form $\Phi = \Phi_{sc} \times
\Phi_{(\alpha, \beta,\gamma)}$, for which $M = \left( \begin{matrix}
\alpha & \beta \\ \beta & \alpha \end{matrix} \right)$.  Tables
\ref{fluxNoTable23} and \ref{fluxNoTable45} lists some examples. 

\section{Summary}

In this paper, we generalized the pattern-of-zeros characterization and
classification of FQH states to multicomponent cases.  We found that
the topological orders in a multicomponent FQH state can be
characterized by the following data: a matrix $n$ that describes the
cluster structure, a matrix $m$ and a sequence $\{S_{\vec a}\}$ that
describes the pattern of zeros.

Our pattern-of-zeros characterization gives us a general quantitative
view on a large class of Abelian and non-Abelian bilayer FQH states,
which allow us to determine which states are simpler than other
states. We find some simplest non-Abelian generalizations of the
Laughlin-Halperin Abelian bilayer states.  Those simple non-Abelian
states may describe some of the bilayer/spin-unpolarized FQH states
observed in experiments and numerical calculations.

This research is partially supported by NSF Grant No.  DMR-0706078.

\appendix

\section{Occupation Number Characterization}
\label{occupationNumberSection}

In the single-component pattern-of-zeros description, there is an occupation number characterization
that is a useful way to understand both the ground states and the quasiparticles in FQH states. 
The generalization to multilayer states does not appear to be quite as simple or useful, but for
the sake of completeness we will analyze it below. 

A convenient set of single-particle basis states for particles in the lowest Landau level are 
monomials of the form $z^m$, for integer $m$. Thus, a basis for symmetrized wavefunctions of
$N_I$ particles of type $I$ is given by:
\begin{equation}
\label{occupationBasis}
\Phi_{ \{\vec{n_l} \}} = \sum_{\{P_I\}} \prod_I \prod_{i=1}^{N_I} (z_{P_I(i)}^I)^{l^I_i},
\end{equation}
where $P_I$ is a permutation of the particles of type $I$, $l^I_i$ is an integer, and 
$\vec{n}_l$ is a vector whose component $n_l^I$ is the number of particles of type $I$
that occupy the $l^\text{th}$ orbit. The polynomial $\Phi$ that 
we are interested in can be expanded in terms of these basis states as:
\begin{equation}
\label{basisExpansion}
\Phi = \sum_{ \{\vec n_l\} } C_{ \{ \vec n_l \}} \Phi_{\{ \vec n_l \}}.
\end{equation} 

Now we may ask what kind of boson occupations $\{ \vec n_l\}$ will be present in the sum (\ref{basisExpansion})
for a polynomial $\Phi$ with a given pattern-of-zeros $\{S_{\vec a}\}$. To answer this question, let us set
$z_1^1 = 0$ in $\Phi(\{ z_i^I\})$. Since $\Phi$ is nonzero when $z_1^1 = 0$ due to translation invariance,
there must be a boson occupation $\{\vec n_l\}$ in the above sum that contains at least one boson occupying
the $(z^1)^{l = 0}$ orbital. That is, there is a term in the above sum with $n_0^1 > 0$. Now, suppose that we
bring a second particle of the same type, $z_2^1$ to 0. The minimal power of $z_2^1$ in $\Phi(0, z_2^1, \cdots)$
is $D_{\vec e_1, \vec e_1}$:
\begin{align}
\Phi(0, z_2^1,z_3^1, ..., z_{N_M}^M) \sim &(z_2^1)^{D_{\vec e_1, \vec e_1}} P_2(z_3^1, z_4^1, ...) 
\nonumber \\
& + O((z_2^1)^{D_{\vec e_1, \vec e_1} + 1}).
\end{align}
Thus, among those $\{ \vec n_l\}$ that have at least one boson of type $1$ occupying the $(z^1)^{l = 0}$
orbital, there must also be an $\{ \vec n_l\}$ that contains a second boson of type $1$ occupying 
the $(z^1)^{l_{2 \vec e_1}}$ orbital where $l_{2 \vec e_1} = D_{\vec e_1, \vec e_1} = S_{2 \vec e_1} - S_{\vec e_1}$.  Next, 
assume that two bosons occupy the $(z^1)^0$ and $(z^1)^{l_{2 \vec e_1}}$ orbital, and bring a third particle
of type $1$ to 0; the minimal power of $z_3^1$ is $D_{2 \vec e_1, \vec e_1}$:
\begin{align}
P_2(z_3^1, z_4^1, ...) \sim &(z_3^1)^{D_{2 \vec e_1, \vec e_1}} P_3(z_4^1, z_5^1, ...)
\nonumber \\
& + O((z_3^1)^{D_{2 \vec e_1, \vec e_1} + 1}).
\end{align}
Thus, among those $\{\vec n_l\}$ that have two type $1$ bosons occupying the $l = 0$ and $l_{2 \vec e_1}$ orbitals,
there is a third boson of type $1$ occupying the $l_{3 \vec e_1} = D_{2 \vec e_1, \vec e_1} = S_{3 \vec e_1} - S_{2 \vec e_1}$ 
orbital. Continuing in this way, we see that there must be a type $1$ boson occupying the orbitals
$l^1_{a \vec e_1} = S_{a \vec e_1} - S_{(a-1)\vec e_1}$ for $a = 1, \cdots, N_1$. After taking all the type $1$
particles to 0, we may begin to take the type 2 particles to zero, one by one, thus obtaining that
there must be a type $2$ boson occupying the orbitals $l^2_{a\vec e_2 + N_1 \vec e_1} = S_{a \vec e_2 + N_1 \vec e_1} - S_{(a-1) \vec e_2 + N_1 \vec e_1}$.
Continuing this argument for bosons of every type, we find that there must be a term in the sum
(\ref{basisExpansion}) with occupation number described by the above sequence of $l^I_{\vec a}$'s. 

However, in the above argument, we chose a particular sequence in which to take various particles to zero. We first took 
all of the type 1 particles to zero one-by-one, and then all of the type 2 ones, and so on. But we could just as
well have made the argument with any sequence. Suppose that after taking $i$ particles to the origin,
there is a $\vec a_i$-cluster at the origin. Thus $\{ \vec a_i\}$ is a sequence that describes the order in
which we take particles to the origin until all particles are at the origin. For every such sequence, 
we may make the above argument and argue that if $\vec a_{i + 1} = \vec a_i + \vec e_I$, then 
there must be a type $I$ boson occupying the orbital $l^I_{\vec a_{i+1}} = S_{\vec a_i + \vec e_I} - S_{\vec a_i}$.  
If we enumerate all the different sequences $\{ \vec a_i \}$ by an integer $\alpha$, then by considering each
$\alpha$, we see that there must be a term in the sum (\ref{basisExpansion}) with occupation number
$\vec n_l^\alpha$. $n_l^{I;\alpha}$ would be the number of $i$, along the sequence $\{ \vec a_i \}$, for which
$l_i^I = l$. Notice that $l^I_{\vec a}$ must be non-zero. Thus we have the following important condition
on $S_{\vec a}$:
\begin{equation}
l^I_{\vec a} = S_{\vec a} - S_{\vec a - \vec e_I} \geq 0.
\end{equation}

The analysis above can be thought of in the following way. Consider an $N_f$-dimensional 
lattice $\mathbb{Z}^{N_f}$, where $N_f$ is the number of layers. At every site $\vec a$ of this
lattice ($a_I \geq 0$) we can associate the nonnegative integer $S_{\vec a}$. On each link 
($\vec a$, $\vec a - \vec e_I$), of the lattice we may also associate an integer 
$l^I_{\vec a} = S_{\vec a} - S_{\vec a - \vec e_I}$. Now consider any directed
path from the origin to $\vec N$ ($N_I$ is the number of particles of type $I$),
in which the sum of the coordinates of every point on the path is one larger than the sum of the
coordinates of the point preceding it, and enumerate the set of these paths by $\alpha$.
To each such path we associate an occupation number sequence
$\vec n_l^{\alpha}$, where $n_l^{I;\alpha}$ is the number of links along the path $\alpha$ 
whose $l^I_{\vec a} = l$. If $\Phi$ has a pattern-of-zeros $\{S_{\vec a}\}$, then its basis expansion
(\ref{basisExpansion}) must contain a term with occupation number 
$\vec n_l^\alpha$. Thus we may rewrite (\ref{basisExpansion}) as 
\begin{equation}
\Phi = \sum_{\alpha} C_\alpha \Phi_{\vec n_l^\alpha} + \sum_{\{ \vec{\tilde n}_l \}} D_{\{ \vec{\tilde n}_l \}} \Phi_{(\{ \vec{\tilde n}_l \})} (\{z_i^I\}). 
\end{equation}
The two sequences $\{S_{\vec a}\}$ and $\{ \vec n_l^\alpha \}$ contain the same information and are one-to-one
labellings of each other. However, $\{ \vec n_l^\alpha \}$ is redundant in the sense that it
does not need to be specified for every $\alpha$ in order to reconstruct $\vec l_{\vec a}$.
The $\vec{ \tilde n}_l$ that appear in the second sum characterize the subleading terms that
appear when coordinates are brought together; thus those $\{\vec{\tilde n}_l\}$ correspond to 
sequences $\{ \tilde{S}_{\vec a} \}$ where $\tilde{S}_{\vec a} \geq S_{\vec a}$.  

In the single-layer case, $\{ l_{a} \}$ naturally defined an occupation number sequence $\{ n_l \}$,
which also described the FQH state in the thin-cylinder limit. In the multi-component generalization,
we have $\{ \vec l _{\vec a} \}$, which seems to admit no simple generalization of the above
occupation number sequence. Instead, one has such occupation number 
distributions for a large number of sequences which we enumerated above by $\alpha$. 
We have not analyzed on general grounds which particular sequences contribute the most
weight to the wavefunction in the thin cylinder limit. 

\section{$su(3)_2/u(1)^2$ Parafermion Conformal Field Theory}
\label{su3para}

Some of the simplest non-Abelian bilayer states are closely related to 
$su(3)_2/u(1)^2$ parafermion CFT. This CFT has central charge $c = 6/5$ and
has three simple currents, $\psi_{\alpha}$, $\psi_{\beta}$, and 
$\psi_{\alpha + \beta} = \psi_{\alpha} \psi_{\beta}$, all of which square 
to the identity and have scaling dimension $1/2$. 

There are four other primary fields, which are associated 
with the fundamental representation of $su(3)$. Their scaling dimensions are
listed in Table \ref{su3paraf}. The fusion rules for these fields all follow from the following
fusion rule:
\begin{equation}
\sigma \times \sigma = 1 + \sigma_{\beta}.
\end{equation}

\begin{table}
\begin{tabular}{|c|c|}
\hline
\multicolumn{2}{|c|} {Primary fields in $su(3)_2/u(1)^2$} \\
\hline
CFT Operator & Scaling Dimension \\
\hline
$\sigma$ & $1/10$ \\
\hline
$\sigma_{\alpha} = \psi_{\alpha} \sigma$ & $1/10$\\
\hline
$\sigma_{\beta} = \psi_{\beta} \sigma $& $6/10$ \\
\hline
$\sigma_{\alpha + \beta} = \psi_{\alpha + \beta} \sigma$ & $1/10$ \\ 
\hline
\end{tabular}
\caption{\label{su3paraf}
Primary fields and their scaling dimensions in the  $su(3)_2/u(1)^2$ parafermion CFT.}
\end{table}

\section{Calculations for candidate states}
\label{calculations}

\subsection{Number of Edge Modes}

The total number of edge modes is equal to the central charge of the corresponding
CFT for the states that are described by the pattern of zeros. For the hierarchy states,
the number of edge modes is given by the rank of the $K$-matrix. 
Furthermore, in the latter case, the number of right (left) -moving
edge modes is given by the number positive (negative) eigenvalues of the $K$-matrix. 

The interlayer Pfaffian states are described by a CFT that consists
of the Ising CFT, with $c = 1/2$, and two scalar boson CFTs, each with $c = 1$. Thus the
number of edge modes for the interlayer Pfaffian is $2\frac{1}{2}$. 

The intralayer Pfaffian states have two Ising CFTs in addition to the two scalar boson
CFTs, so the total number of edge modes is $3$. 

The central charge of the $su(3)_2/u(1)^2$ parafermion CFT is $c = 6/5$; the two-component FQH states
based on this are described by the $su(3)_2/u(1)^2$ theory and two scalar bosons, for a total
of $2\frac{6}{5}$ edge modes. 

For the $(m,m,l)$ states, the $K$-matrix is $K = \left( \begin{matrix} m & l \\ l & m \end{matrix} \right)$.
These states have 2 edge modes; if $m > l$, all edge modes move in the same direction; 
if $m < l$, then there is one right-moving and one left-moving edge mode. 

For the states $(\nu_0, \nu_0|m)$, the $K$ matrix is
\begin{equation}
K = \left( \begin{matrix} 
K^0_{11} & K^0_{12} & m & 0 \\
K^0_{21} & K^0_{22} & 0 & 0 \\
m & 0 & K^0_{11} & K^0_{12} \\
0 & 0 & K^0_{21} & K^0_{22} \\
\end{matrix} \right),
\end{equation}
where $K^0$ is the $K$-matrix in the hierarchical basis of
the state $\Phi_{\nu_0}$. For the $\nu = 2/5$ state, 
$K^0$ in the hierarchical basis is 
\begin{equation}
K^0 = \left( \begin{matrix}
3 & -1 \\
-1 & 2 \\
\end{matrix} \right) .
\end{equation}
For the $\nu = 2/7$ state, $K^0$ in the hierarchical basis is
\begin{equation}
K^0 = \left( \begin{matrix}
3 & 1 \\
1 & -2 \\
\end{matrix} \right) .
\end{equation}
For the $\nu=2/3$ P-H conjugate of the $1/3$ Laughlin state, $K^0$
in the hierarchical basis is
\begin{equation}
K^0 = \left( \begin{matrix}
1 & 1 \\
1 & -2 \\
\end{matrix} \right) .
\end{equation}
Using this, we find that the P-H conjugate of the $1/3$ Laughlin state 
has edge modes $1_R + 1_L$, $(2/3,2/3|1)$ has edge modes $2_R + 2_L$, and
$(2/3, 2/3|2)$ has $1_R + 3_L$. 

\subsection{Shifts on Sphere}

In the hierarchy basis, the formula for the shift is given by\cite{WZ9253}
\begin{equation}
\mathcal{S} = \frac{1}{\nu} \sum_I (K^{-1})_{1I} K_{II} .
\end{equation} 
Using this formula, we find $\mathcal{S} = 0$ for the particle-hole
conjugate of the $\nu=1/3$ Laughlin state.  

Now consider the bilayer composite fermion state $(\nu_0, \nu_0|m)$:
\begin{equation}
\Phi_{(\nu_0, \nu_0|m)} = \prod_{i,j} (z_i - w_j)^m 
\Phi_{\nu_0}(\{z_i\}) \Phi_{\nu_0} (\{w_i\}).
\end{equation}
Let $N_{\Phi}^0$ be the maximum power of $z_1$ in $\Phi_{\nu_0}(\{z_i\})$.
It satisfies:
\begin{equation}
N_{\Phi}^0 = \nu_0^{-1} N_1 - \mathcal{S}_0,
\end{equation}
where $\mathcal{S}_0$ is the shift of the state $\Phi_{\nu_0}$. The factor
$\prod_{i,j} (z_i - w_j)^m$ increases the power of $z_1$ by $m N_2$. 
Thus the maximum power of $z_1$ in $\Phi_{(\nu_0, \nu_0|m)}$ is $N_{\Phi}^1$:
\begin{equation}
N_{\Phi}^1 = \nu_0^{-1} N_1 + m N_2 - \mathcal{S}_0. 
\end{equation}
In our cases, $N_1 = N_2$, and the number of flux quanta is the same in each
layer, so
\begin{equation}
N_{\Phi}^1 = (\nu_0^{-1} + m) N_1 - \mathcal{S}_0,
\end{equation}
$\nu_1^{-1} = \nu_0^{-1} + m$ is the filling fraction in one layer. Thus
we see that the shift of $\Phi_{(\nu_0, \nu_0|m)}$ is also $\mathcal{S}_0$. 

For $(2/3,2/3|m)$, we take $\Phi_{2/3}$ to be the particle-hole conjugate
of the $1/3$ Laughlin state. $\Phi_{2/3}$ has shift $\mathcal{S}_0 = 0$.
Thus $(2/3,2/3|m)$ also has shift $\mathcal{S} = 0$. $(2/5,2/5|m)$ has
shift $4$, because $\Phi_{2/5}$ has shift $4$.

For the Halperin $(m,n,l)$ states, the $K$-matrix can be written as
$K = \left( \begin{matrix} m & l \\ l & n \end{matrix} \right)$. In this
basis, the shift is given by
\begin{equation}
\mathcal{S} = \nu^{-1}\sum_{IJ}K^{-1}_{IJ} K_{JJ}.
\end{equation}

For the states described by the pattern of zeros, we can use the following formula 
\begin{equation}
\label{shift}
\mathcal{S} = \left\{
  \begin{array}{ll}
    \nu^{-1} \sum_{I} \nu_I ( m_{II} - S_{\vec n_I} + S_{\vec n_I - \vec e_I})  & \mbox{ for bosons} \\
    \nu^{-1} \sum_{I} \nu_I ( m_{II} + 1 - S_{\vec n_I} + S_{\vec n_I - \vec e_I}) & \mbox{ for fermions } \\
  \end{array}\right.
\end{equation}

\subsection{Electron and Quasiparticle Operators for $su(3)_2/u(1)^2$ States}

The electron operators for the $su(3)_2/u(1)^2$ FQH states that we discuss are of the form:
\begin{align}
V_{e1} &= \psi_{\alpha} e^{i \sqrt{\frac{1}{\nu}} \phi_+ + i s \phi_-},
\nonumber \\
V_{e2} &= \psi_{\beta} e^{i \sqrt{\frac{1}{\nu}} \phi_+ - i s \phi_-},
\end{align}
where $s = \sqrt{3}/2$, $1/2$, and $\sqrt{7}/2$ for the $\nu = 4/5$, $4/7$ and $4/9$ states,
respectively. The quasiparticle operators with minimal total charge are of the form:
\begin{equation}
V_{qp} = \sigma e^{i Q \sqrt{\frac{1}{\nu}} \phi_+ + i s_{qp} s \phi_-},
\end{equation}
and have scaling dimension $h_{qp} = \frac{1}{10} + \frac{Q^2}{2\nu} + \frac{(s_{qp} s)^2}{2}$.
The total charge of the quasiparticle is $Q = 1/5$, $1/7$, and $1/9$
for  the $\nu = 4/5$, $4/7$ and $4/9$ states, respectively.  
$s_{qp} = 1/3$, $1$, and $1/7$, respectively, for these states. 


\end{document}